\documentclass[
10pt,
showpacs,preprintnumbers,nofootinbib,
 amsmath,amssymb,
 aps,
prl,twocolumn,groupedaddress,superscriptaddress,
floatfix,
showkeys
]{revtex4-1}

\usepackage[dvips]{graphicx}
\usepackage{latexsym,epsfig,bm,psfrag}
\usepackage{color}         
\usepackage{dcolumn}                 
\usepackage[bookmarksnumbered,bookmarksopen,colorlinks,citecolor=blue,linkcolor=blue]{hyperref} %
\usepackage{calrsfs}
\usepackage{aas_macros}
\usepackage{float}

{}  
\DeclareMathOperator{\pr}{pr} 
\newcommand{\given}{\,|\,}  

\begin{document}

\renewcommand{\vec}{\boldsymbol}
\newcommand{\mc}{M_{\mathrm{c}}}
\newcommand{\mn}{M_{\mathrm{n}}}
\newcommand{\mnc}{M_{\mathrm{nc}}} 
\newcommand{\mr}{M_{_{\mathrm{R}}}}
\renewcommand{\S}[1]{S_{#1}}
\renewcommand{\P}[1]{P_{#1}}
\newcommand{\D}[1]{D_{#1}}
\newcommand{\half}{\frac{1}{2}}
\newcommand{\threehalf}{\frac{3}{2}}
\newcommand{\Gaf}[1]{\Gamma\!\left(#1\right)}
\newcommand{\eq}[1]{Eq.~(\ref{#1})}
\newcommand{\eqp}[1]{Equation~(\ref{#1})}
\newcommand{\eqtoeq}[2]{Eqs.~(\ref{#1})--(\ref{#2})}
\newcommand{\eqtoeqp}[2]{Equations~(\ref{#1})--(\ref{#2})}
\newcommand{\pe}{p}
\newcommand{\MH}{M_H}
\newcommand{\ith}{i^{\textrm{th}}}
\newcommand{\jth}{j^{\textrm{th}}}
\newcommand{\lth}{\ell^{\textrm{th}}}
\newcommand{\first}{$1^{\textrm{st}}$}
\newcommand{\second}{$2^{\textrm{nd}}$}
\newcommand{\abi}{{\it ab initio \/}}
\newcommand{\nx}{N_\mathrm{max} }
\newcommand{\ex}{e_\mathrm{max} }
\newcommand{\nbar}{n_{\Lambda}}
\newcommand{\nE}{n_{E}}
\newcommand{\omg}{\omega}
\newcommand{\omgT}{\omega_T}
\newcommand{\ghgf}[3]{_{#1}F_{#2}\left(#3 \right)}
\newcommand{\Eexact}{\vec{E}}
\newcommand{\Eabi}{\vec{E}_\star}
\newcommand{\gere}{\mathcal{E}}
\newcommand{\ufunc}{\mathcal{U}}
\newcommand{\uvc}{\Lambda_\mathrm{UV}}
\newcommand{\nuipara}{\vec{\theta}}
\newcommand{\he}[1]{^{#1}\mathrm{He}}
\newcommand{\oxy}[1]{^{#1}\mathrm{O}}
\newcommand{\CL}{\vec{C}_{\mathrm{L}}}
\newcommand{\CH}{\vec{C}_{\mathrm{H}}}
\newcommand{\geretrun}{N_\mathcal{O}}
\newcommand{\gereT}{\overline{\mathcal{E}}}
\newcommand{\qref}{Q_{\mathrm{ref}}}
\newcommand{\xilin}[1]{\textcolor{red}{#1}}

\title{{\it Ab initio} calculations of low-energy nuclear scattering\texorpdfstring{\\}{}  using confining potential traps}

%

\author{Xilin Zhang}
\email{zhang.10038@osu.edu}
\affiliation{Department of Physics, The Ohio State University, Columbus, Ohio 43210, USA}
\author{S.~R.~Stroberg}
\email{stroberg@uw.edu}
\affiliation{Physics Department, University of Washington, Seattle, WA 98195, USA}
\author{P.~Navr\'{a}til}
\email{navratil@triumf.ca}
\affiliation{TRIUMF, 4004 Wesbrook Mall, Vancouver, British Columbia, V6T 2A3 Canada}
\author{Chan Gwak}
\affiliation{TRIUMF, 4004 Wesbrook Mall, Vancouver, British Columbia, V6T 2A3 Canada}
\affiliation{Department of Physics and Astronomy, University of British Columbia, 
Vancouver, 
BC
V6T 1Z1, Canada
} 
\author{J.~A.~Melendez}
\affiliation{Department of Physics, The Ohio State University, Columbus, Ohio 43210, USA}
\author{R.~J.~Furnstahl }
\affiliation{Department of Physics, The Ohio State University, Columbus, Ohio 43210, USA}
\author{J.~D.~Holt}
\affiliation{TRIUMF, 4004 Wesbrook Mall, Vancouver, British Columbia, V6T 2A3 Canada}
\affiliation{Department of Physics, McGill University, 3600 Rue University, Montr\'eal, QC H3A 2T8, Canada}

\date{August, 2020\\[20pt]}

\begin{abstract}
A recently modified method to enable low-energy nuclear scattering results to be extracted from the 
discrete energy levels of the target-projectile clusters confined by harmonic potential traps is tested. 
We report encouraging results for neutron--$\alpha$ and neutron--$^{24}\mathrm{O}$ elastic scattering from analyzing the trapped levels computed using two different \abi nuclear structure methods.
The $n$--$\alpha$ results have also been checked against a direct \abi reaction calculation. The $n$--$^{24}\mathrm{O}$ results demonstrate the approach's applicability for a large range of systems provided their spectra in traps can be computed by \abi methods. 
A key ingredient
 is a rigorous understanding of the errors in the calculated energy levels  caused by inevitable Hilbert-space truncations in the \abi methods. 
\end{abstract}

\maketitle

\paragraph{Introduction} Low-energy nuclear theory has entered an era of precision calculations, thanks to the systematic development of nuclear forces~\cite{Epelbaum:2008ga}, \abi many-body methods~\cite{Barbieri:2016uib,Barrett:2013nh,Carlson:2014vla,Lee:2008fa,Hagen:2013nca,Stroberg:2019mxo} that use these interactions as input, and uncertainty quantification~\cite{Melendez:2017phj,Melendez:2019izc,Zhang:2015ajn,Zhang:2019odg}. This enables rigorous computer simulations of nature and provides a tool for studying nuclear systems that is complementary to real experiments. 
It also improves the nuclear physics input that is vital to astrophysics, particle physics, and other domains.  

\emph{Ab initio} nuclear scattering/reactions calculations, however, are still limited to a small set of systems~\cite{Nollett:2006su,Navratil:2016ycn,Elhatisari:2015iga,Shirokov:2018nlj}, while structure has progressed to medium-mass and even heavy nuclei~\cite{Barbieri:2016uib,Lee:2008fa,Stroberg:2019mxo,Hagen:2013nca,Morr18Tin,Mane20Cd}. A compelling strategy~\cite{Luu:2010hw,Rotureau:2011vf,Zhang:2019cai} is to expand the former's reach by taking advantage of the latter's progress: use  structure methods to compute discrete energy levels for projectile-target (p-t) clusters in harmonic potential traps, and then extract free-space scattering/reaction observables from the levels. Recently, Ref.~\cite{Zhang:2019cai} has improved the method to allow systematic control of theory errors. 

The approach is similar in spirit to the L{\"u}scher method used in  Lattice QCD~\cite{Luscher:1990ux}, which extracts hadronic scattering observables from energy spectra discretized by a spatial box with periodic boundary conditions. 
Both the trap and box \emph{physically} reduce the number of degrees of freedom (dofs) to enable spectrum calculations. 
Note the trap preserves rotational invariance (broken for anisotropic traps) and the decoupling between internal and center-of-mass (CM) dynamics. 

The modified trap formula~\cite{Zhang:2019cai}  is in fact a quantization condition (QC) in the nonrelativistic limit for p-t \emph{relative} dynamics in two-cluster elastic scattering: given angular momentum $\ell$ and trap frequency $\omgT$, the  eigenenergy $E$ satisfies a transcendental equation of the form
\begin{equation}
   \gere_\ell (\omgT,E)  = \ufunc_{\infty,\ell} (\omgT, E) \; . \label{eqn:master1}
\end{equation} 
Here, a nucleon (with mass $M_N$) at location $\vec{r}_i$ experiences a potential $\frac{1}{2} M_N \omgT^2 \vec{r}_i^2$; for p-t separated by $\vec{r}$, the relative potential (with reduced mass $\mr$) is $\frac{1}{2} \mr \omgT^2 \vec{r}^2$. (The static form of the potential fixes our reference frame.) When the relative momentum $p \equiv \sqrt{2\mr E}$  and the inverse of the trap length scale $\sqrt{\mr \omgT} \equiv 1/b_T$ are smaller than the high momentum (UV) scale $\MH$, the p-t's internal dofs can be integrated out and the left side can be expanded in terms of $b_T^{-4}$ and $p^2$ with coefficients $C_{i,j}$:
\begin{equation}
 \gere_\ell    =\sum_{i,j=0}^{\infty}  C_{i,j}\times {\left(\mr \omgT \right)^{2i} p^{2j}}  \; .    \label{eqn:Edef}  
\end{equation} 
This equation generalizes the conventional effective range expansion (ERE); namely the phase shift  $\delta_\ell(E)$ in partial wave $\ell$ is obtained from the ERE~\cite{vanKolck:1998bw,Hammer:2017tjm}
\begin{equation}
   p^{2\ell+1}\cot\delta_\ell\left(E\right) =  \sum_{j=0}^{\infty} C_{i=0,j}\times p^{2j} \;. 
\end{equation}
The terms with $C_{i\neq 0,j}$ account for how the trap modifies the p-t interaction at short distance~\cite{Zhang:2019cai}. 
Dimensional estimates suggest that $C_{i,j} \sim \MH^{2\ell+1-4i-2j}$. 
The right side, called a ``unitarity function'' here, has its analytic structure dictated by long-distance (IR) physics, irrespective of the UV physics:
\begin{equation}
  \ufunc_{\infty,\ell}  = (-)^{\ell+1}
  \left(\frac{2}{b_T} \right)^{ 2 \ell +  1 }\, 
  \frac{\Gaf{- \nE}}{\Gaf{-\ell-1/2 - \nE }} \; ,  
  \label{eqn:Udef}   
\end{equation} 
with $\nE \equiv E/\!\left(2\omgT\right) -\ell/2 -3/4 $. 
When $\omg_T\! \rightarrow\! 0$, $ \mathcal{U}_{\infty,\ell} \rightarrow i p^{2\ell +1}$ and the infinite poles of $\ufunc_{\infty,\ell}$ in the complex $E$ plane coalesce into the usual unitarity branch cut. 
Further details about \eqtoeq{eqn:master1}{eqn:Udef} and previous works~\cite{Luu:2010hw,Rotureau:2011vf,Busch1998,PhysRevA.65.043613,PhysRevA.65.052102,PhysRevA.66.013403,PhysRevLett.96.013201,Stetcu:2007ms,PhysRevA.80.033601,Stetcu:2010xq,Rotureau:2010uz,Blume_2012} can be found in Ref.~\cite{Zhang:2019cai}.

If the \abi  methods we use---specifically, the no-core shell model (NCSM)~\cite{Barrett:2013nh} and the valence-space formulation of the in-medium similarity renormalization group (VS-IMSRG)~\cite{Stroberg:2019mxo}---used an infinite Hilbert space, then \eqtoeq{eqn:master1}{eqn:Udef}  would apply directly.
However, in practice the Hilbert space is truncated, modifying both IR and UV physics. 
These methods construct their many-body Hilbert spaces using a single-particle basis of harmonic oscillator (HO) wave functions, with basis frequency $\omg$.
The NCSM limits the system's total HO excitation quanta (relative to the naive level filling) to be below $\nx$, whereas the IMSRG assigns a cutoff $\ex$ to each nucleon; both act as UV \emph{and} IR regulators. 

To model the regulator-induced errors, we study a two-body problem: 
the {\it relative} Hamiltonian using the same HO basis with a cutoff $\nbar$ on the  {\it radial} excitation quanta. 
The unitarity function now depends on $\nbar$ and $\omg$:
\begin{eqnarray}
   \ufunc_{\ell}&& (\nbar,\omg;\,\omgT, E) = \notag \\ 
   && ( - )^{\ell+1}\! \left(\frac{2}{x\, b_T} \right)^{\! 2 \ell\!+\!  1 } 
    \frac{\Gaf{{3}/{2}\!+\! \ell}}{\Gaf{{1}/{2}\! - \! \ell}  } 
    \frac{\Gaf{\nbar\!+ 2 }}{\Gaf{\nbar\!+\ell +  {5}/{2} }  } \notag \\ 
 && \times 
  \frac{\ghgf{2}{1}{n_{_E} \!+ 1,\; -\nbar \! - \ell  - {3}/{2};\; 1/2 - \ell;\; x^2 } }{ \ghgf{2}{1}{n_{_E} \!+\ell+3/2,\; -\nbar\! -  1;\; 3/2+ \ell;\;  x^2   } } \; , \label{eqn:Utrundef}   
\end{eqnarray}
with $b \equiv 1/\sqrt{\mr \omg} $, $x \equiv 2 b_T b / (b_T^2 + b^2 )$, and $\ghgf{2}{1}{, ; ;}$ defined in~\cite[Eq.~16.2.1]{NIST:DLMF}. 
Note that $\ufunc_{\ell}(\nbar,\omg;\omgT, E) \rightarrow \ufunc_{\infty,\ell} (\omgT, E) $ with either $\nbar \rightarrow \infty$ (no space-truncation) or $\omg \rightarrow \omgT$ (basis has correct IR physics).
Moreover, for integer $\nbar$, $\ufunc_{\ell}$ has $\nbar + 1$ poles located at the eigenvalues of the truncated HO Hamiltonian.
%

The  QC is now 
\begin{equation}
\gere_\ell (\uvc;\omgT,E) = \ufunc_{\ell}(\nbar,\omg;\omgT, E)  \; , \label{eqn:master2}    
\end{equation} 
with $\gere_\ell$ depending on the 
regulator-induced 
UV-cutoff scales ($\uvc$); i.e., $C_{i,j} \rightarrow C_{i,j}(\uvc)$ in \eq{eqn:Edef}. 
Because the IR-modification is fully accounted for in $\ufunc_\ell$, the  error of the extracted $\gere_\ell (\uvc;\omgT,E)$ and $C_{i,j}(\uvc)$ via \eq{eqn:master2} is UV in nature and reduces to zero when $\uvc$ 
is greater than the UV scale of the nucleon interaction. 
The QC \eqref{eqn:master2} and its $\uvc$ dependence is explicated in the Supplemental Material (SM)~\cite{Zhang:2020rhzSM}.  

In this paper, we use $n$--$\alpha$ and $n$--$^{24}\mathrm{O}$ scattering as examples to show that the UV and IR errors in the \abi eigenenergy outputs can be modeled through \eq{eqn:master2}. 
The former serves as a benchmark, by comparison to  results from an existing direct \abi calculation (using no-core shell model with continuum (NCSMC)~\cite{Navratil:2016ycn}), while the latter demonstrates that the approach is applicable in larger systems where no \abi treatment exists to date. 
The derivation of the new QC and the details in analyzing the \abi output will be presented in  two subsequent papers \cite{Zhang20203,Zhang20202}.

\paragraph{{\it Ab initio} calculations} 
Both methods use the chiral effective field theory nucleon  interaction NNLO${\rm opt}$~\cite{PhysRevLett.110.192502}, which provides a good description of light nuclei including the oxygen isotopes. We do not apply any renormalization of the interaction.
The NCSM extracts the low-energy eigenvalues and eigenstates numerically through matrix diagonalization. The  $\nx$ cutoff guarantees the factorization of the CM wave function from the intrinsic wave function~\cite{Barrett:2013nh}. To directly compute scattering/reactions, the clustering states with correct asymptotic behavior of the inter-cluster wave function are included in the Hilbert space (this approach is known as NCSMC). 

The  IMSRG~\cite{Tsuk12SM,Bogn14SM,Hergert:2015awm,Stro17ENO,Stroberg:2019mxo}  applies  unitary transformations~\cite{Morr15Magnus} to the Hamiltonian to decouple the low- and high-energy Hilbert spaces, which produces an effective low-energy Hamiltonian. 
The impact of induced many-body operators are assumed to be small and are therefore neglected here.
This assumption has been validated in numerous benchmark calculations, e.g.~\cite{Hergert2013,Parzuchowski2017EM}.
Unlike the $\nx$ cutoff, the IMSRG's $\ex$ cutoff  couples CM with internal dofs, but this coupling is reduced with increasing $\ex$ and has been demonstrated to be minimal for converged calculations~\cite{Hagen:2009pq,Hergert:2015awm}.  
A rigorous estimation of these two types of errors in the IMSRG is left for future study. 
In the following, all the \abi energies have the CM energy subtracted (also for NCSM). 

The computational resources needed for the NCSM and NCSMC grow exponentially with the number of nucleons, while  for the IMSRG they grow polynomially. 
Therefore only the latter is currently feasible for calculations of medium-mass nuclei. 
Both the NCSM and IMSRG are well-suited for computing self-bound nuclei.
Trapping nucleons with theoretically imposed external fields makes scattering systems artificially bound, and thus requires little modification to these \abi methods.
The trapping interaction, which is proportional to $\vec{r}_i^2$, can be analytically expressed in the HO basis and thus including it is straightforward.

To extract the phase shifts in a given partial wave, the quantum numbers of the p-t system need to match those of the individual p and t (e.g., for $n$--$\alpha$  $\P{3/2}$ scattering, the computed states are the $\alpha$ ground state ($0^+$) and $\he{5}$ $3/2^-$ state in various traps). Here, we only use the lowest p-t eigenenergy within a given channel; other states corresponding to radial excitation could be useful and will be explored in the future. The \abi output and information about all the computed states can be found in the SM~\cite{Zhang:2020rhzSM}.

\begin{figure*}
  \includegraphics{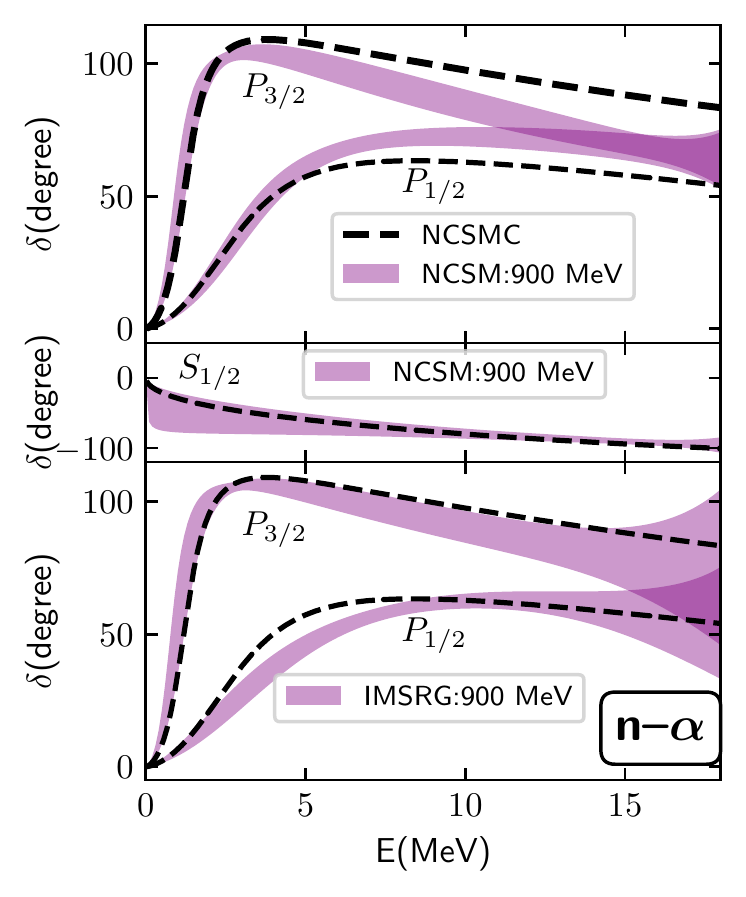}
 \hspace{-0.15in}
  \includegraphics{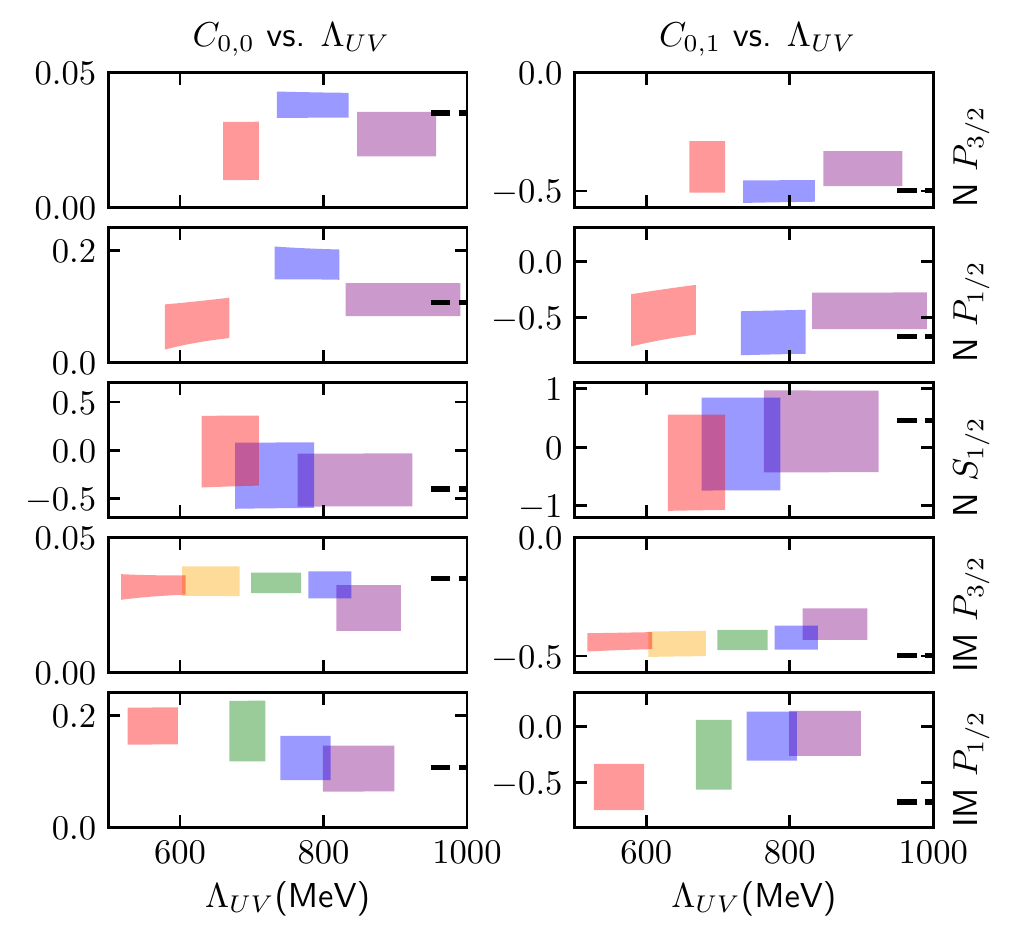}
  \caption{Results for $n$--$\alpha$ scattering. The right panel: $C_{0,0}(\uvc)$ and $C_{0,1}(\uvc)$ vs $\uvc$. The rows are for different \abi methods (``N'' for NCSM and ``IM'' for IMSRG) and scattering channels. Each error band comes from analyzing individual binned data sets (see the SM~\cite{Zhang:2020rhzSM} for detailed information).   The left panel plots error bands for scattering phase shifts from different \abi methods and channels. The dashed lines in both panels are from direct NCSMC calculations.} \label{fig:nHe4}  
\end{figure*}

\paragraph{Data analysis} 
We label $\alpha$ or $\oxy{24}$ as t and the neutron as p. 
The right side of \eq{eqn:master2} evaluated at the p-t relative eigenenergies from the \abi calculations
is equated to the generalized ERE (GERE)  expansion from Eq.~\eqref{eqn:Edef}. 
(The expansion's convergence radius is $E\leq E_H \equiv \MH^2/2\mr = 22$ and 4\,MeV for $n$--$\alpha$ and $n$--$\oxy{24}$ scatterings, as determined by the targets' lowest excitation energies.)
To get the relative eigenenergy, the \abi p-t total energy ($E_{pt}$) must have the associated t energy ($E_t$) subtracted. The HO basis ensures that $E_{pt}$ and $E_t$ should come from the same trap and regulators with the same $\omg$, but \textit{a priori} $\nx^t \neq \nx^{pt}$ or $\ex^t \neq \ex^{pt}$. Thus we have 
%
\begin{eqnarray}
  E(\nx^{pt}, \omg, \omgT) =
  E_{pt}(\nx^{pt}, \omg, \omgT)- E_{t}(\nx^{t}, \omg, \omgT). 
\label{eqn:deltatidlenmaxdef}
\end{eqnarray}
We proceed by linearizing the difference between $\nx^t $ and $\nx^{pt}$ in \eq{eqn:deltatidlenmaxdef} about $\omgT$ for each set of $\nx^{pt}$ and $\omg$.  
Since $E_{t}$ is only known for integer $\nx^{t}$, we interpolate those points  with given $\omg$ and $\omgT$ to get smooth functions.  

The relation between $\nbar$ used in $\ufunc_{\ell}$ and the many-body regulator $\nx^{pt}$  is also \textit{a priori} unknown (though $\omg$ and $\omgT$ should be the same). 
Again, we parameterize the difference between $2\nbar+ \ell$ and $\nx^{pt}$ as a linear function of $\omgT$ for given $\nx^{pt}$ and $\omg$,  
%
%
and allow $\nbar$ to be a non-integer.
%
%
For the IMSRG analysis, the same linear models are employed for inferring $E$ from  $E_{pt}$ and $E_t$, and $\nbar$ from $\ex^{pt}$ and $\omg$. The details  are provided in the SM~\cite{Zhang:2020rhzSM}. 
The model parameters are now collectively labelled as  $\nuipara$.

The data sets are then binned based on their estimated $\uvc$ values. 
We take $\uvc = \sqrt{(2\ex + 7)M_N\omg} $ for the $\ex$ regulator, considering that the largest eigenvalue of the single-nucleon momentum-squared operator  $\vec{p}^2$ in the  truncated  Hilbert space is  $\uvc^2$~\cite{Binder:2015trg,Konig:2014hma}. For the $\nx$ regulator, $\sqrt{(2\nx + 7) M_N\omg} $ (for $\alpha$ and $\he{5}$) has the same meaning if $\nx$ quanta are assigned to a single nucleon, which is used here as a nominal $\uvc$ for this regulator. $\nx$ ($\ex$) takes the value of $\nx^{t}$ ($\ex^{t}$) in \eq{eqn:deltatidlenmaxdef}. This $\uvc$  represents the UV-cutoff scales for the targets ($\alpha$ and $\oxy{24}$), and should also be positively correlated with the UV-cutoff scales for the relative motion.

Each data bin has a $\uvc$ width on the order of $100$\,MeV, across which we expect only mild changes of $C_{i,j}$.  
Therefore a simple interpolation formula should suffice:
\begin{equation}
 C_{i,j}(\uvc)=\sum_{k=0}^{1}  C_{i,j,k} \times (\qref/\uvc)^{2k} \ , \label{eqn:lambdaUVintp}
\end{equation}
 with $\qref$ as a parameter. 

In short, our  error sources include the error induced by truncating GERE's series in \eq{eqn:Edef}, and those---\emph{modeled} using $\nuipara$ and $\qref$---caused by truncating the many-body Hilbert-space and the poor understanding of its impacts on subsystems and relative motion. Future study uncovering the nature of $\nuipara$'  will reduce the latter error. 

Here, we rely on data analysis to constrain $\nuipara$, $\qref$, and $C_{i,j,k}$, by using Bayesian inference~\cite{Sivia1996, Furnstahl:2014xsa,Furnstahl:2015rha}. 
Each $\uvc$  bin is analyzed independently to produce 
errors for the observables in this bin. The details are provided in the SM~\cite{Zhang:2020rhzSM} and  Ref.~\cite{Zhang20202}. In the following results, the first error becomes significant when $E \rightarrow E_H$, while the second type dominates in other regions. (The numerical errors in the computed eigenenergies are rounding errors and much smaller than these two.)

\begin{figure}[t]
  \includegraphics[width=\columnwidth]{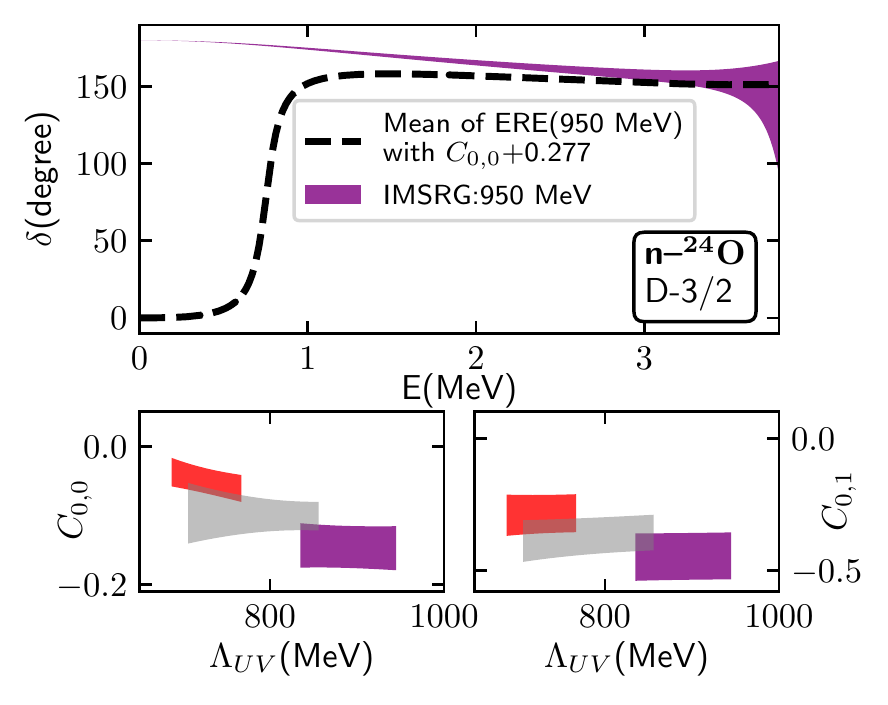}
  \caption{Results for $n$--$\oxy{24}$ scattering. The top panel shows the error band of the extracted phase shift. The dashed line corresponad to particular values of $C_{i,j,k}$ as detailed in the text. The bottom panel shows $C_{0,0} (\uvc)$ and $C_{0,1} (\uvc)$ against $\uvc$ extracted from analyzing three different binned data sets (c.f.\ SM~\cite{Zhang:2020rhzSM} for detailed information). } \label{fig:nO24}
\end{figure}

\paragraph{Results}
Figure~\ref{fig:nHe4} shows the $n$--$\alpha$ results. The right panel plots the 1-$\sigma$ error bands (vertical axis) for  $C_{0,0}$ and $C_{0,1}$ against $\uvc$ for various channels and from different \abi methods. 
Here and below, $C_{i,j,k}$ is rescaled by $\MH^{2\ell+1-4i-2j}$  and becomes dimensionless.
Each block is from analyzing one binned data set, whose regulator parameters can be found in the SM~\cite{Zhang:2020rhzSM}. 
The width of $\uvc$ is determined by its distribution among the binned data points (the region with $68\%$ degrees of belief). 
Note that the correlation between $C_{0,0}$ and $C_{0,1}$'s errors is nonzero, although not  shown here. The black dashed lines mark the NCSMC results. A na\"ive estimate suggests $C_{i,j} \sim 1$, but $C_{0,0}$ is constrained to be $\sim 10^{-2}$ ($10^{-1}$) in the $\P{3/2}$ ($\P{1/2}$) channels from both \abi calculations, while the other parameter values are consistent with the estimate. Other terms not shown here, such as $C_{0,2}$ and $C_{1,0}$, are also well constrained to non-zero values~\cite{Zhang20202}.  

It is worth highlighting the smoothness in the $C_{i,j}$'s $\uvc$ dependence, given that the bins are extracted from different regulators. This signals that the regulator-induced IR-error is properly modeled; otherwise the $\ufunc_\ell$'s $E$ dependence near its poles, as controlled by $\nbar$, could induce non-smooth behavior in $C_{i,j}(\uvc)$. 
An illustrative example from the two-body model is provided in the SM~\cite{Zhang:2020rhzSM}. 
Also note that the $\uvc$ scales in the NCSM and IMSRG results are not easily connected; thus their $\uvc$ dependencies could be different. 

Figure~\ref{fig:nHe4}'s left panel shows the phase shift error bands as transformed from the 1-$\sigma$  bands of $\gere_\ell(\uvc;\omgT=0,E)$. It has two contributions added in quadrature: one due to the uncertainty in $C_{i,j}(\uvc)$ and the other from truncating the GERE series expansion, as detailed in the SM~\cite{Zhang:2020rhzSM}.  
$\uvc$ is set at 900\,MeV, where $C_{i,j}$ apparently converges. 
The agreement between the NCSM phase shifts and the dashed lines (NCSMC) at low energy is not only a benchmark for our method but also a self-consistent check of the NCSMC calculation. 
The disagreement at higher energy is not understood at present, but might stem from the  modeling of the $\Delta$ and $\tilde{\Delta}$ functions. Note the NCSMC phase-shift uncertainty was estimated to be about 5\% (see SM~\cite{Zhang:2020rhzSM}).

The IMSRG phase shifts are similar to the NCSMC phase shifts in $\P{3/2}$, while in $\P{1/2}$ they differ at low energy. This could be due to the truncation of many-body operators (the spin-orbit splitting between the two channels is sensitive to three-body forces~\cite{Nollett:2006su,Hupin:2013wsa}). 
Note the error bands' rapid increase with $E\rightarrow 20$\,MeV is due to the GERE-series-truncation error, 
which diverges outside the theory's applicability region.  Since the $^{5}$He system is treated in a p-shell valence space in the IMSRG calculation, the method cannot access the $\S{1/2}$ channel. 

In Fig.~\ref{fig:nO24}, the analogous results are provided for $n$--$^{24}\mathrm{O}$ ($\D{3/2}$). We only use $\ex=14$ data in the analysis (see the SM~\cite{Zhang:2020rhzSM} for details), which limits the number of bins shown here. 
Again, a clear but smooth $\uvc$ dependence emerges for $C_{i,j}$. 
We compute a 1-$\sigma$ band for $\gere(\uvc; \omgT=0, E)$ with $\uvc=950$\,MeV and transform it to the phase shift band in the left panel. 
Existing experimental information~\cite{Caesar:2012tva,Jones:2017lrs} indicates a resonance at 0.75\,MeV with a width about 90 keV, while our extracted phase shift indicates the existence (with $75\%$ probability) of a shallow bound state with binding energy at $-1.4 \pm 0.5\,$MeV.

Note that $C_{0,0}(\uvc)$ increases with decreasing $\uvc$ while $C_{0,1}$ is more stable, hinting at a positive $C_{0,0}$ and thus a low-energy resonance at $\uvc < 600\,$MeV. 
This demonstrates that modifying the nucleon interaction (here through changing the regulator) could reproduce a resonance. 
We illustrate this by applying the mean value of $C_{i,j}(\uvc=950\,\mathrm{MeV})$ and increasing $C_{0,0}$ by 0.277 in $\gere(\uvc; \omgT=0, E)$, producing the dashed curve in the top panel. 
It indicates a resonance at 0.75\,MeV with a 135\,keV width, which is compatible with the experimental information. 
This implies that the nucleon interaction could be tuned to reproduce the resonance. 
It is worth mentioning that our IMSRG calculation using the same nucleon interaction without a trap shows that $^{25}\mathrm{O}$ is unbound against one-neutron separation. 
 In contrast, the system is found to be shallowly bound after the continuum physics is correctly included.  

\paragraph{Summary} 

We have modified the method of confining harmonic traps and implemented it for \abi calculations (NCSM and IMSRG) for $\he{4,5}$ and $\oxy{24,25}$ nuclei. 
We successfully extracted the elastic scattering phase shifts from the ground state energies at various traps and with different regulators. 
For $n$--$\alpha$, the extracted phase shifts from both \abi results are in good agreement with the direct NCSMC calculation within uncertainties. 
For $n$--$^{24}$O, we also extract phase shifts and find it necessary to fine tune the underlying nucleon interaction to reproduce experimental information. 
Our method provides a unified framework to treat continuum physics and shallow bound states, as currently needed in low-energy nuclear physics~\cite{Johnson:2019sps}.

\paragraph{Acknowledgment}  
The work of XZ, JAM, and RJF  was supported in part by the National Science Foundation under Grant Nos. PHY–1614460  and PHY--1913069, and the NUCLEI SciDAC Collaboration under US Department of Energy MSU subcontract RC107839-OSU. TRIUMF  receives  funding  via a  contribution  through  the  National  Research  Council of Canada. 
The work of PN was supported by NSERC grant No. SAPIN-2016-00033 and by an INCITE Award on the Titan supercomputer of the Oak Ridge Leadership Computing Facility (OLCF) at ORNL.
This  work  was  further supported  by NSERC and the Canadian Institute of Nuclear Physics.
SRS was supported by the US
Department of Energy under contract DE-FG02-97ER41014.
Computations were performed with an allocation of computing resources on Cedar at WestGrid and Compute Canada, and  on  the Oak  Cluster  at  TRIUMF  managed  by  the University of British Columbia department of Advanced Research Computing (ARC).
We are grateful to the Institute for Nuclear Theory for support under INT Program INT-19-2a, 
``Nuclear Structure at the Crossroads”. During the program, we made significant
progress on this project.

\bibliographystyle{apsrev4-1}
%

\clearpage
\onecolumngrid
\begin{center}
 \textbf{\large Supplementary Material for {\it Ab initio} calculations of low-energy nuclear scattering using confining potential traps}\\[.4cm]
  Xilin Zhang,$^1$ S.~R.~Stroberg,$^2$ P.~Navr\'{a}til,$^3$ Chan Gwak,$^{3,4}$ J.~A.~Melendez,$^1$ R.~J.~Furnstahl,$^1$ and J.~D.~Holt$^{3,5}$\\[.1cm]
  $^1${\itshape Department of Physics, The Ohio State University, Columbus, Ohio 43210, USA} \\
  $^2${\itshape Physics Department, University of Washington, Seattle, WA 98195, USA} \\ 
  $^3${\itshape TRIUMF, 4004 Wesbrook Mall, Vancouver, British Columbia, V6T 2A3 Canada} \\ 
  $^4${\itshape Department of Physics and Astronomy, University of British Columbia, Vancouver, BC V6T 1Z1, Canada} \\ 
  $^5${\itshape Department of Physics, McGill University, 3600 Rue University, Montr\'eal, QC H3A 2T8, Canada} 
\end{center}

\twocolumngrid

\setcounter{equation}{0}
\setcounter{figure}{0}
\setcounter{table}{0}
\setcounter{section}{0}
\setcounter{page}{1}
\makeatletter
\renewcommand{\theequation}{S\arabic{equation}}
\renewcommand{\thefigure}{S\arabic{figure}}

\maketitle

\section{Additional details}

\begin{figure*}
  \includegraphics{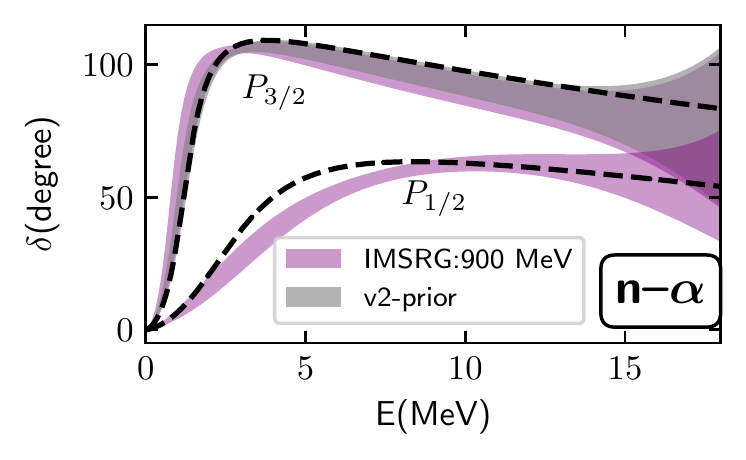}
 \hspace{-0.15in}
  \includegraphics{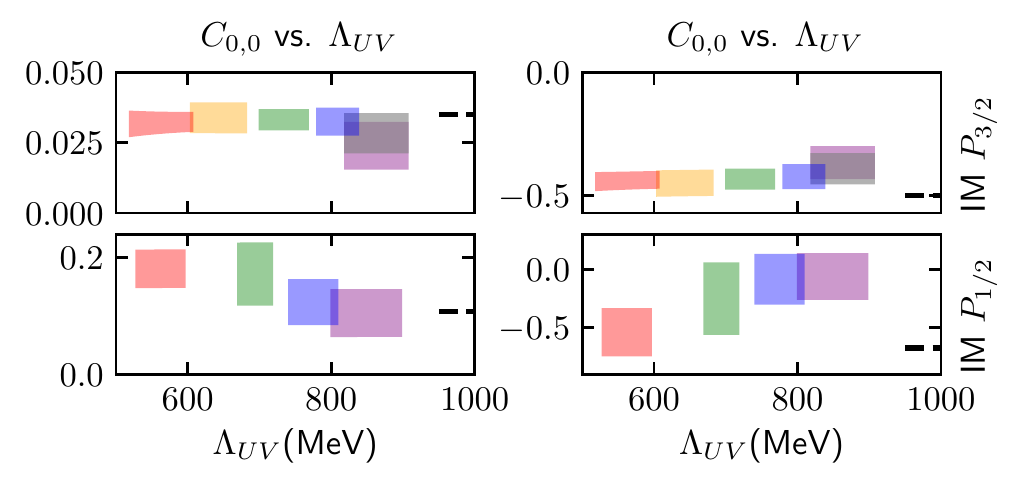}
  \caption{Results for $n$--$\alpha$ scattering based on the IMSRG output. The dashed lines in both panels are from direct NCSMC calculations. This plot is part of Fig.~\ref{fig:nHe4} (main text) with the gray bands included, which are the results of a Bayesian inference using a more restrictive prior.} 
  \label{fig:nHe4_SM}  
\end{figure*}

Here we provide additional details/comments that could be helpful for reading the paper. 

\begin{itemize}
\item The GERE expansion in \eq{eqn:Edef} could have an unnaturally small convergence radius if there are low-energy poles/cuts in the GERE function, which could be caused by fine-tuning or other long-range forces such as the Coulomb force. To increase the convergence radius, those non-analyticities need to be explicitly identified and included. Some discussion on this point can be found in Ref.~\cite{Zhang:2019cai}.

    \item
Note that the eigenenergies used to infer the GERE values  can also be computed using other {\abi} structure methods (e.g., quantum monte-carlo calculation, GFMC~\cite{Carlson:2014vla}), but their output's errors and error-propagation to the extracted phase-shifts need to be studied (without the errors, \eq{eqn:master1} in the main text is also applicable).

   \item
In the right panel of Fig.~\ref{fig:nHe4}, the IMSRG results are binned more finely than the NCSM's, because the former provides more data. 

\item 
  In Fig.~\ref{fig:nHe4}, the error bars in the $\S{1/2}$ channel are larger in general than those in other channels. This is because there are less data collected in this channel than in other channels from NCSM eigenenergy calculations (see Table~\ref{tab:He-NCSM-s1-2}). In this channel, the eigenenergies are generally larger than those in the p-wave channels for the same $\omgT$; a portion of them are even larger than $E_H$ and hence not used in data analysis.

  \item
When discussing the $n$--$\alpha$ phase-shift extractions in the main text, the uncertainty of the direct phase shift calculation by the NCSMC method was mentioned to be about 5\%.  This value was estimated by varying $\omg$ between 20 and 28\,MeV, $\nx$ up to $17$, and the number of $^5$He composite eigenstates up to 8 within the calculation. 
We note that a more rigorous error estimate method for NCSMC is now under development~\cite{Kravvaris:2020lhp}. 
 
  \item
 In Fig.~\ref{fig:nHe4}'s right panel, for the highest $\uvc$ bin in the IMSRG $\P{3/2}$ channel, there are double modes in the parameter fitting. 
After reducing our prior window to exclude the mode with larger size of $\Delta$ and $\tilde{\Delta}$, the $C_{i,j}$ error bands (gray and ``v2-prior'' in Fig.~\ref{fig:nHe4_SM}) are better aligned with neighbouring bins than the original (purple); the new phase shift is also shown in gray in Fig.~\ref{fig:nHe4_SM}.

    \item 
Priorities for going forward:    
To reduce the phase shift error bands, the origin of the nonzero $\Delta$ and $\tilde{\Delta}$ (defined in Eqs.~\ref{eqn:deltatidlenmaxdef2} and~\ref{eqn:deltanmaxdef}) needs to be better understood, perhaps by studying other observables and many-body wave functions. 
To be applied to charged-particle scattering, our modified trap formula needs to include the Coulomb interaction. 
Another important step is studying coupled-channel  reactions and three-body scattering/reactions using the same strategy. 
Note that the parallel topics to these are being actively studied in  Lattice QCD. 
Thus, the studies outlined here could provide valuable cross-field benchmarks for the general strategy, considering that for specific nuclear systems there exists other \abi scattering/reaction methods (e.g., NCMSC and GFMC).

\item It is worth pointing out some differences between our work and previous approaches developing \abi scattering and reaction calculations, such as the so-called SS-HORSE method~\cite{Shirokov:2016thl, Shirokov:2018nlj} and 
the calculations combining the coupled-cluster method and the Gamow Hartree-Fock basis for scattering nucleon~\cite{Hagen:2012rq, Hagen:2013jqa}. In terms of final results, our work provides rigorous error bars for the extracted scattering phase shifts, while the above mentioned works have yet to do so. Both SS-HORSE and our method share the same spirit as that of a  L{\"u}scher-type method. In fact, the unitarity function in the SS-HORSE method is \eq{eqn:Utrundef} with $\omgT=0$ (i.e., without trap). However, the treatment of IR error due to unknown connection between  $\nbar$ and $\nx$ has not been studied in the SS-HORSE works~\cite{Shirokov:2016thl, Shirokov:2018nlj}. In addition, our method can be implemented with other structure methods such as GFMC as already mentioned above, while SS-HORSE can only be implemented with the structure methods using harmonic oscillator wave function as basis.  
\end{itemize}

\section{A two-body model}
Here we apply our approach to a two-body model, which was constructed in Ref.~\cite{Ali:1984ds} to qualitatively reproduce $n$--$\alpha$ scattering phase-shifts in its s and p waves. 
This model was also used in Ref.~\cite{Zhang:2019cai} to study \eq{eqn:master1}, i.e.,  the modified trap method without accounting for errors in the input eigenenergies. 
The potential between the two particles take the form of a square well with spin-orbital interactions:  $V_s(r) =  V_0 (1+ \beta \vec{L}\cdot\vec{\sigma})$ when $r < r_c$ and $0$ when $r > r_c$, with $V_0=-33\,$MeV, $r_c=2.55\,$fm, and $\beta=0.103$~\cite{Ali:1984ds}.

To test the application of the modified trap formula in \eq{eqn:Utrundef}, 
we first compute the exact (untruncated) energy spectrum with two different $\omgT$ values, and rely on \eq{eqn:Udef} to compute $\gere_\ell(\omgT,\Eexact)$ at these \emph{exact} eigenenergies $\Eexact$. 
These discrete points are then interpolated to form a continuous function, approximating the full $\gere_\ell(\omgT, E)$ function [labeled as $\gere^\mathrm{exact}_\ell(\omgT, E )$]. 
We then construct the Hamiltonian using a truncated HO basis and compute its eigenenergies $\Eabi$ 
for various truncations of the relative motion excitation quanta $\nbar$. Plugging $\Eabi$ and the corresponding regulator parameter values into \eq{eqn:Utrundef}, we can reconstruct $\gere_\ell(\omgT,\Eabi)$ (labeled as $\gere^\mathrm{regulated}_\ell(\omgT,\Eabi)$). 
In general $\Eabi \neq \Eexact$, so in the following results, we choose $\omgT$ values that represent those used in the \abi calculations in the main text, but also make sure $\Eabi$ is close to $\Eexact$. 
For both p-wave channels we use $\omgT=2$ and $10$\,MeV, but for the s-wave channel we use $\omgT=0.5$ and $10$\,MeV 
($\omgT=0.5$ is chosen to have the eigenenergies
$\Eexact$ 
closely separated to minimize interpolation errors).  

\begin{figure}[tbh]
  \includegraphics[width=0.45 \textwidth]{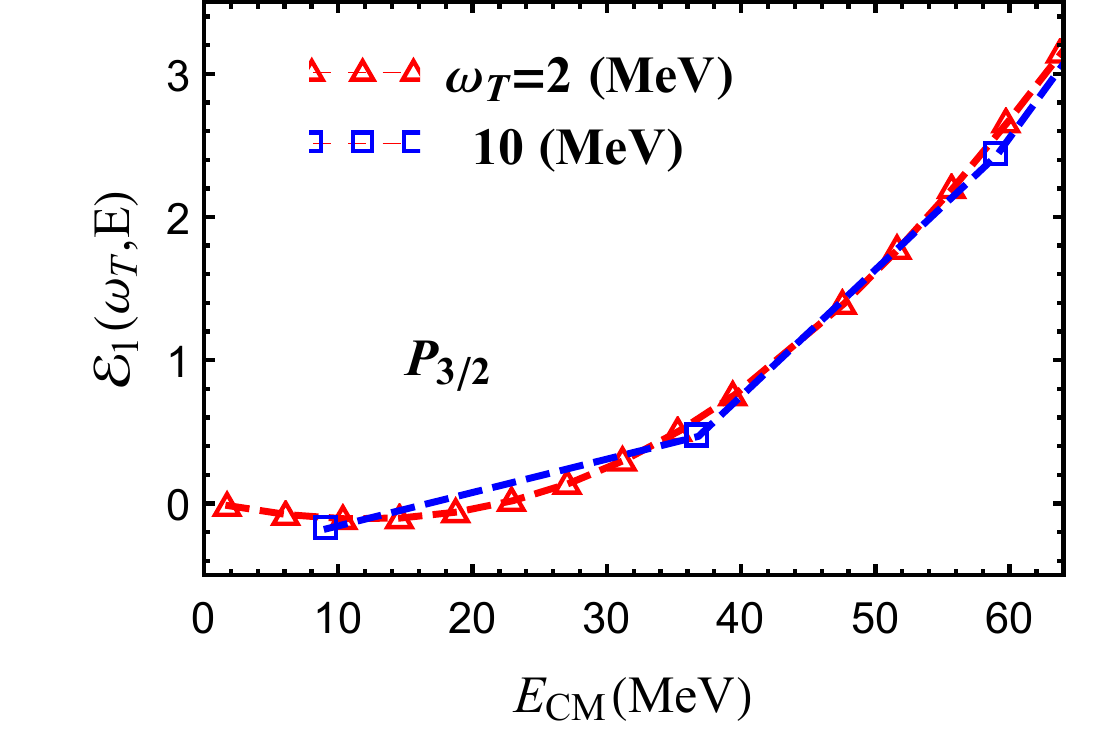}
  \caption{ $\gere^\mathrm{exact}_\ell(\omgT, \Eexact)$ values at the exact eigenenergies (i.e., without Hilbert-space truncation) for the $\P{3/2}$ channel. } \label{fig:toyp3halfgere}
\end{figure}
 
\begin{figure}[tbh]
  \includegraphics[width=0.45 \textwidth]{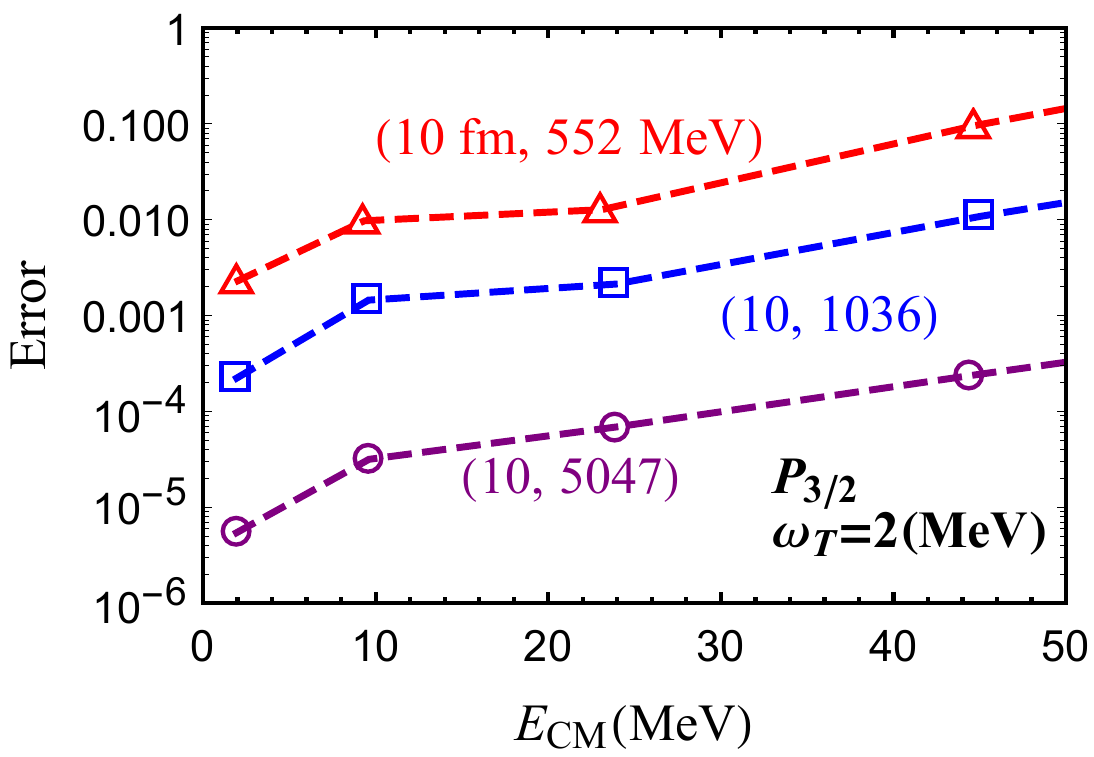}
  \includegraphics[width=0.45 \textwidth]{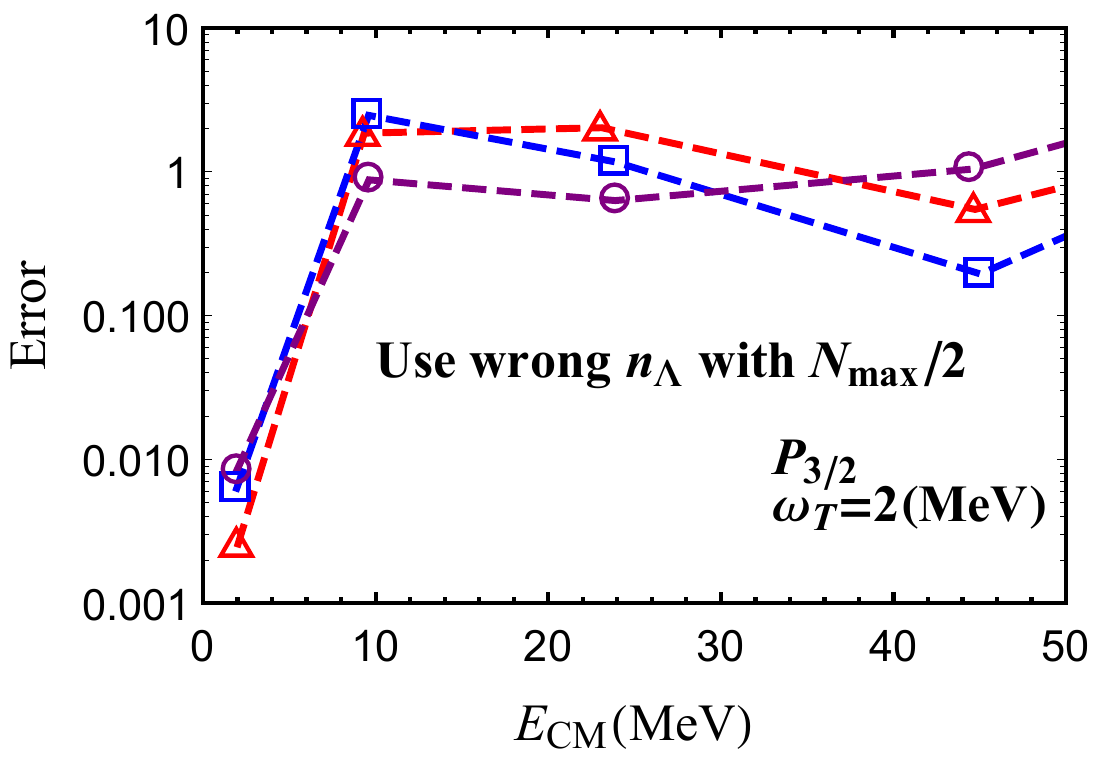}
    \includegraphics[width=0.45 \textwidth]{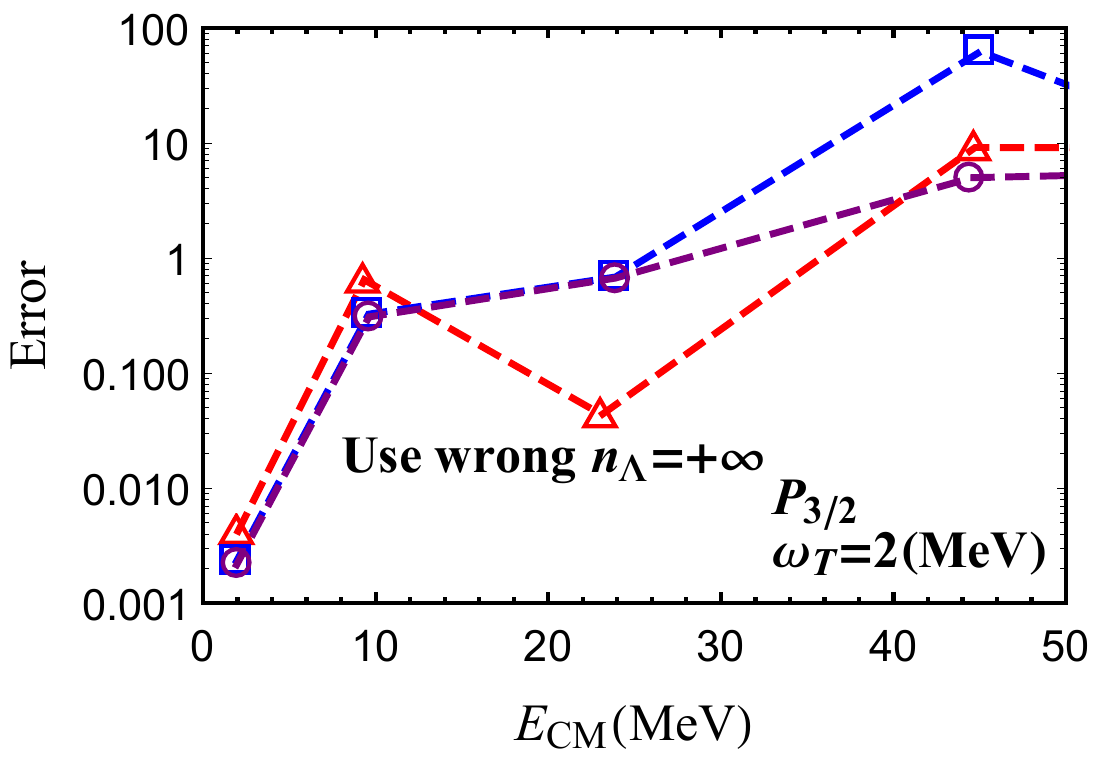}
  \caption{The absolute error of extracted $\gere^\mathrm{regulated}_\ell(\omgT=2\,\mathrm{MeV},\Eabi)$ values (i.e., its difference from the exact values $\gere^\mathrm{exact}_\ell(\omgT,\Eabi)$) at eigenenergies of the truncated Hamiltonian for the $\P{3/2}$ channel. 
  The plot labels [e.g., (10\,fm, 552\,MeV)] refer to the $L_\mathrm{IR}$ and $\uvc$ values of the used regulators. 
  The $(\nx ,\omgT)$ values for these calculations are $(11,14)$, $(23,27)$, $(125,132)$, which are ordered by increasing $\uvc$. 
  The top panel uses the correct $\nbar$, while the lower two panels use incorrect $\nbar$ values, as noted in the plots. 
  See the text for the details. } \label{fig:toyp3half}
\end{figure}

\begin{figure}[tbh]
  \includegraphics[width=0.45 \textwidth]{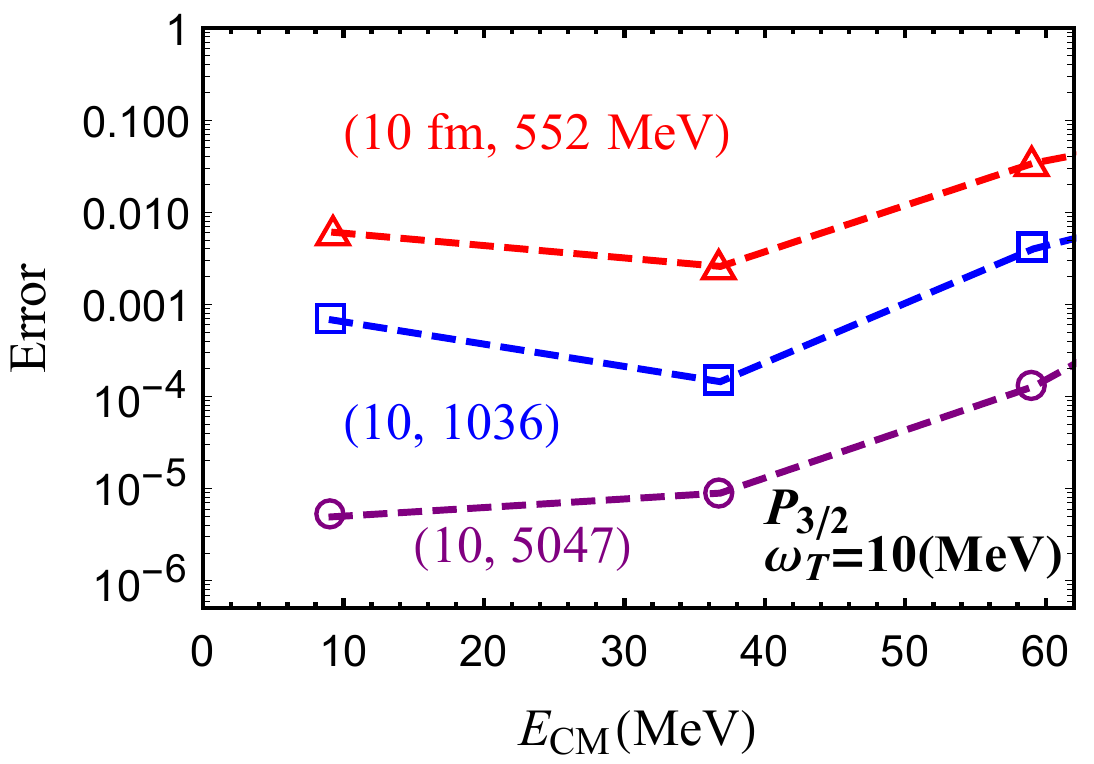}
  \includegraphics[width=0.45 \textwidth]{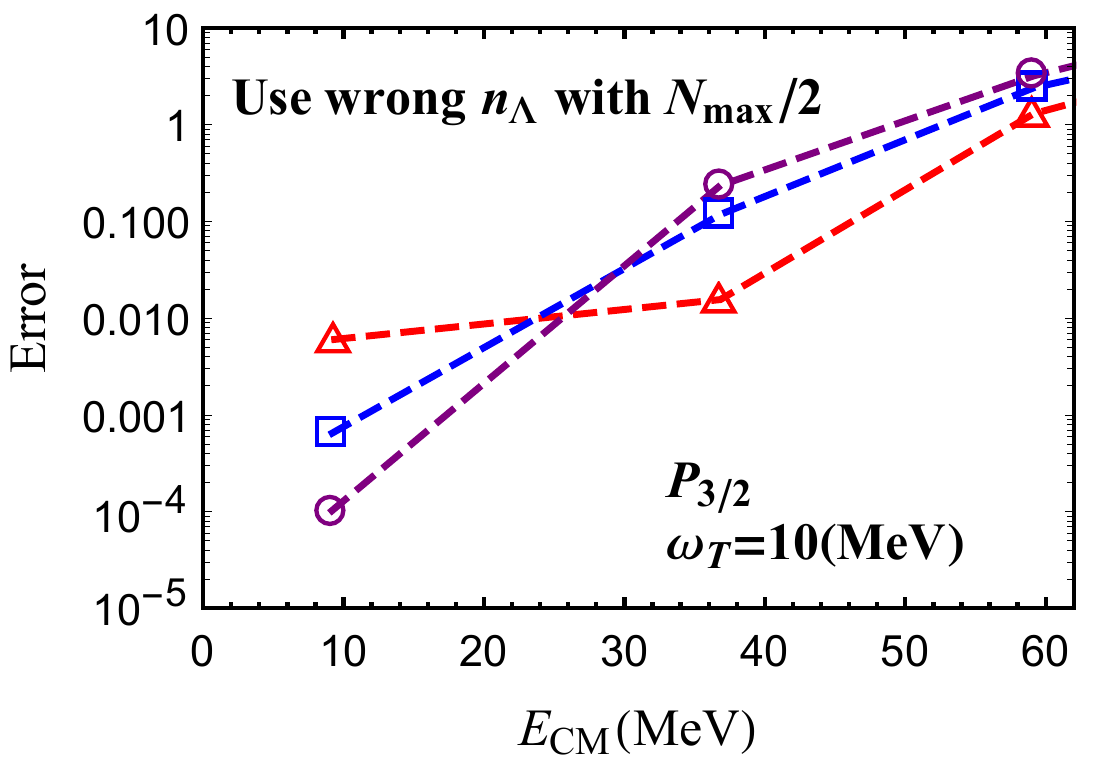}
   \includegraphics[width=0.45 \textwidth]{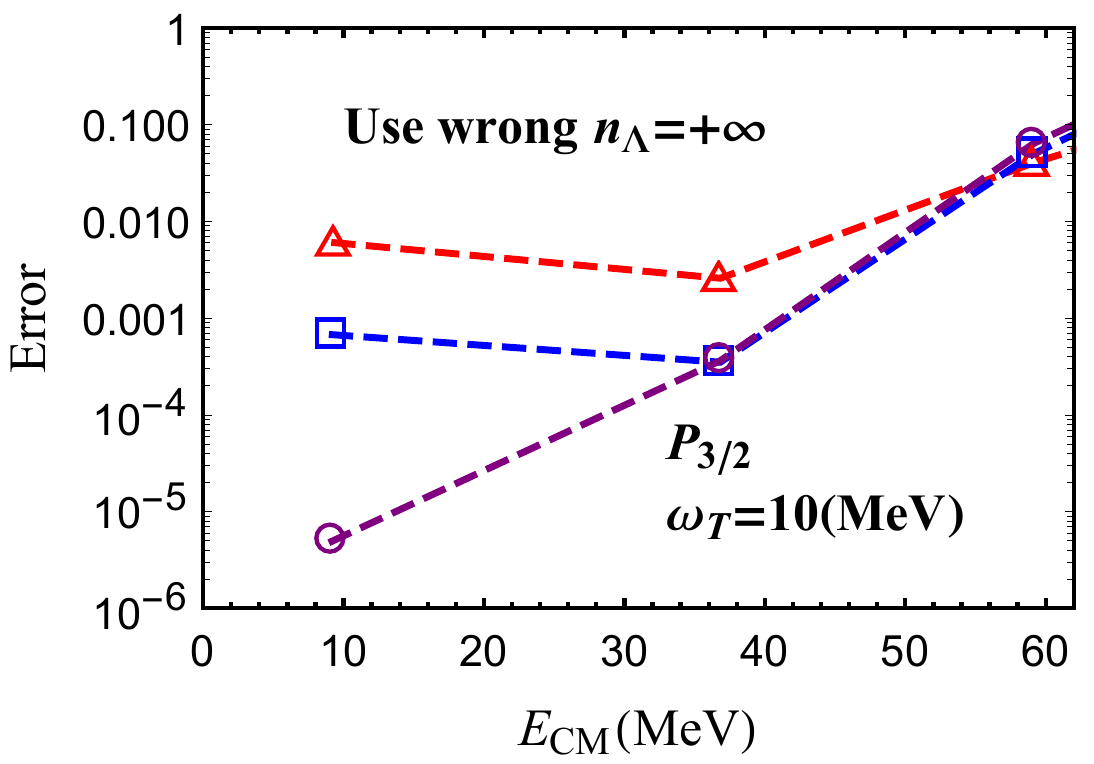}
  \caption{
   Parallel plots for $\omgT=10$\,MeV as those in Fig.~\ref{fig:toyp3half} for $\P{3/2}$ channel. See the captions there for the $(\nx ,\omgT)$ values.  } \label{fig:toyp3half2}
\end{figure}

Note that in the current section the $\gere_\ell$ and $\ufunc_\ell$ values are re-scaled by a reference scale $M_\mathrm{ref}^{2\ell + 1}$ ($M_\mathrm{ref} = 200\,$MeV is chosen to be the same as that used in Ref.~\cite{Zhang:2019cai}). 
Also note that Ref.~\cite{Zhang:2019cai} shows the high-energy scales in this model for all the channels are in the range of between $20$ and $50\,$MeV. So in the following plots, we show figures up to $E\sim 50$\,MeV.

Figure~\ref{fig:toyp3halfgere} shows $\gere^\mathrm{exact}_\ell(\omgT, \Eexact )$ for the  p-wave ($3/2^-$) at $\Eexact$ from the corresponding exact calculation without Hilbert-space truncation.  
The $\gere^\mathrm{exact}_1(\omgT, E )$ function varies from being on the order of  $10^{-2}$ to  $10^{-1}$ when $E$ is below $10\,$MeV to being on the order of $1$ at higher energies. 
 
The top panels in Figs.~\ref{fig:toyp3half} and~\ref{fig:toyp3half2} show  the absolute value of $\gere^\mathrm{regulated}_\ell(\omgT,\Eabi)-\gere^\mathrm{exact}_\ell(\omgT,\Eabi)$ in the  $\P{3/2}$ channel, i.e., the error of reconstructed $\gere_\ell$ values by plugging corresponding $\omgT$, $\Eabi$, $\omg$ and $\nbar$ in \eq{eqn:Utrundef}. 
The labels of the calculations using different $\nbar$ and $\omg$ correspond to the infrared length scale $L_\mathrm{IR} \equiv \sqrt{2\nx + 7}\; b$ (fm) and $\uvc \equiv \sqrt{2\nx + 7}/ b $ (MeV) with $\nx = 2 \nbar + \ell$ and $b\equiv 1/\sqrt{\mr \omg}$. 
Note that $\uvc$ defined here  differs from  the one used in analyzing \abi outputs in the main text by a $\sqrt{\mr/M_n}$ factor ($\mr$ is the $n$--$\alpha$ reduced mass while $M_n$ is the mass of a nucleon). 
The definition of $L_\mathrm{IR}$ is motivated in a similar way: it corresponds to the largest eigenvalue of $\vec{r}^2$ in the truncated relative-motion Hilbert space. 
The values of the corresponding $\nx$ and $\omg$ can be found in  Fig.~\ref{fig:toyp3half}'s caption. 
$L_\mathrm{IR}$ is chosen to be the same in the three calculations,  i.e., the calculations have the same IR conditions.

We clearly see that the error decreases systematically with increasing $\uvc$.
This trend reflects the convergence toward the exact result as $\uvc \rightarrow \infty$ of the reconstructed phase-shift using \eq{eqn:Utrundef} and the eigenenergies of the truncated Hamiltonian. The significant dependence on $\uvc$ also shows the necessity of introducing $\uvc$ dependence in the left side of \eq{eqn:master2}.  
The lower two panels in those figures show the same error by using incorrect $\nbar$ values in \eq{eqn:Utrundef}: the middle panel uses $\nbar$ with $2\nbar+\ell$, which is half of the correct $\nx$, and the bottom panel uses $\nbar \rightarrow \infty$, which is equivalent to using \eq{eqn:Udef}. 
It is clear that if you use the wrong $\nbar$,  the errors could be $100\%$ or even larger (c.f.\ Fig.~\ref{fig:toyp3halfgere}) and the reconstructed $\gere_\ell$ values would not be able to be fitted using a smooth curve with a ``length'' scale on the order of 10\,MeV. 
We should expect the reconstructed $C_{i,j}(\uvc)$ using the wrong 
$\gere^\mathrm{regulated}_\ell$ values would not be smooth between different $\uvc$ (i.e., different regulators). 
In other words, smooth $\uvc$-behavior signals a correct modeling of the IR physics associated with the truncated Hamiltonian. 

\begin{figure}[tbh]
  \includegraphics[width=0.45 \textwidth]{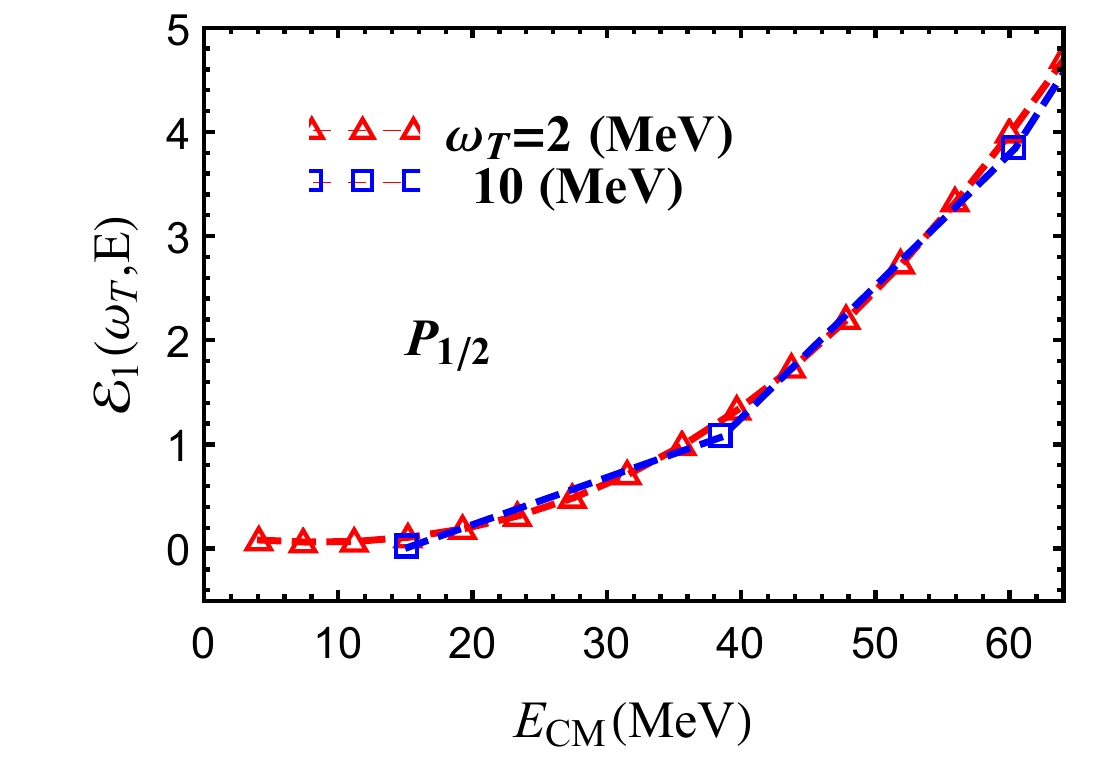}
  \caption{ $\gere^\mathrm{exact}_\ell(\omgT, \Eexact)$ values at exact eigenenergies without Hilbert-space truncation for the $\P{1/2}$ channel. } \label{fig:toyp1halfgere}
\end{figure}

\begin{figure}[tbh]
  \includegraphics[width=0.45 \textwidth]{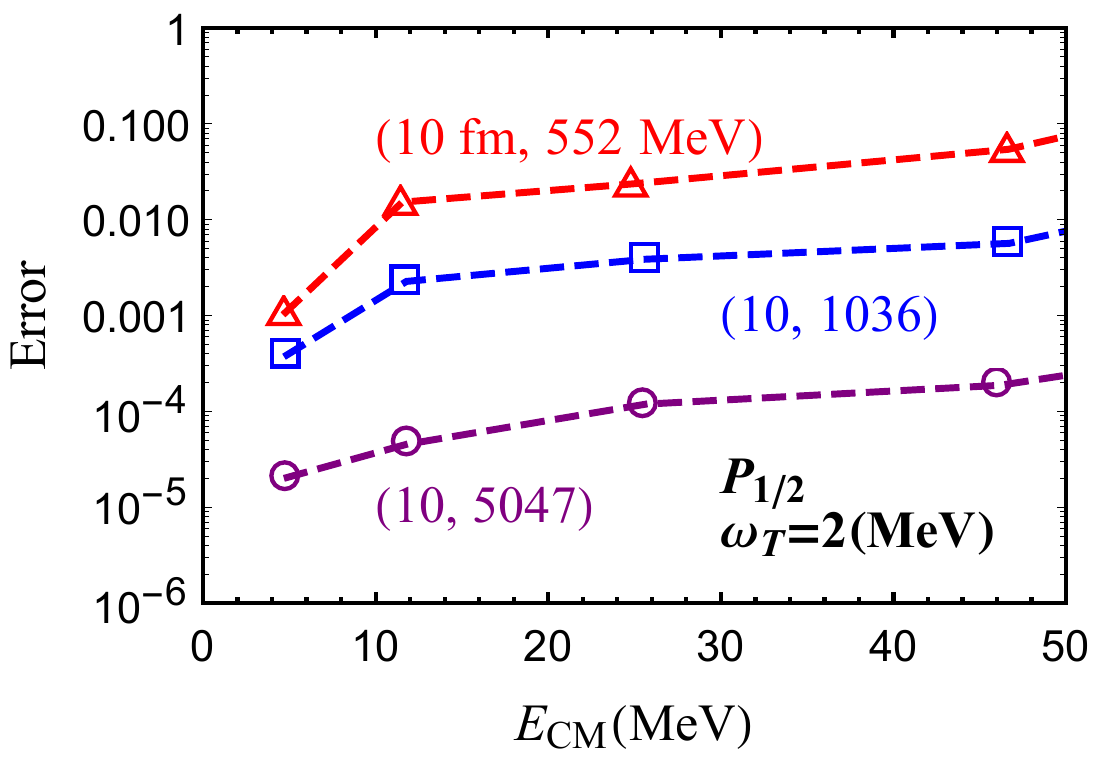}
  \includegraphics[width=0.45 \textwidth]{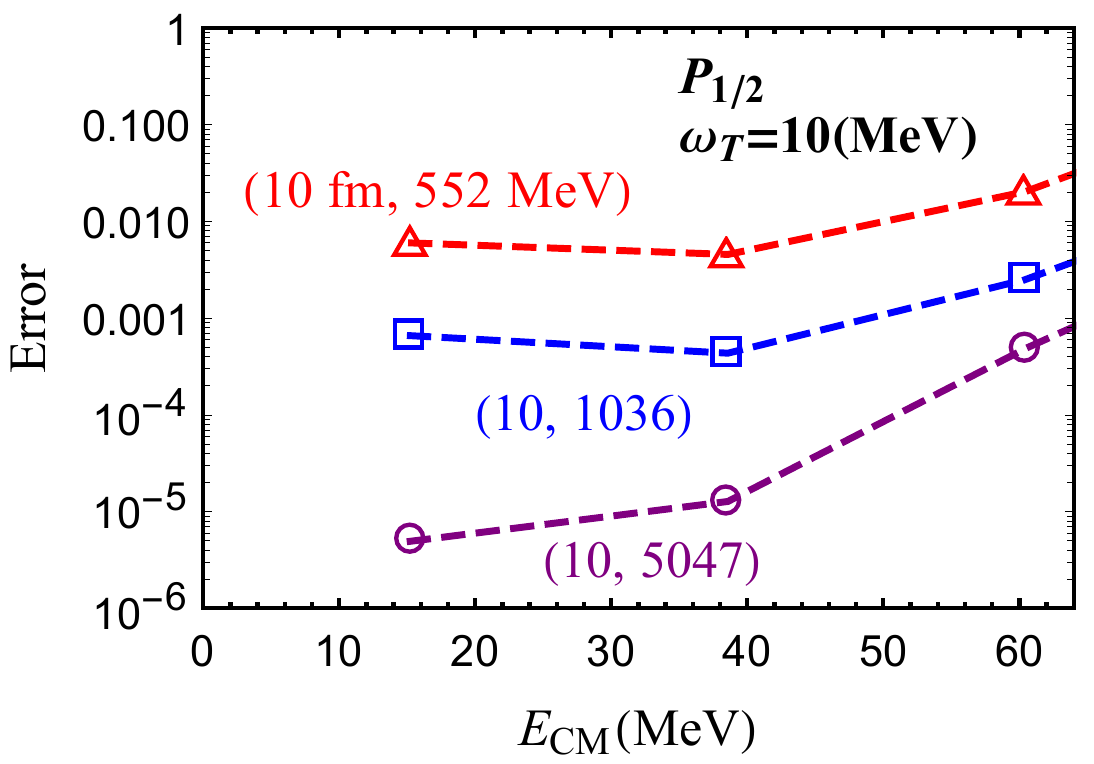}
  \caption{The error of extracted $\gere^\mathrm{regulated}_\ell(\omgT,\Eabi)$ values (i.e., its difference from the exact values $\gere^\mathrm{exact}_\ell(\omgT,\Eabi)$) at eigenenergies of the truncated Hamiltonian for the $\P{1/2}$ channel. The $(\nx ,\omgT)$ values for these calculations are the same as those in $\P{3/2}$ channel.} \label{fig:toyp1half}
\end{figure}

\begin{figure}[tbh]
  \includegraphics[width=0.45 \textwidth]{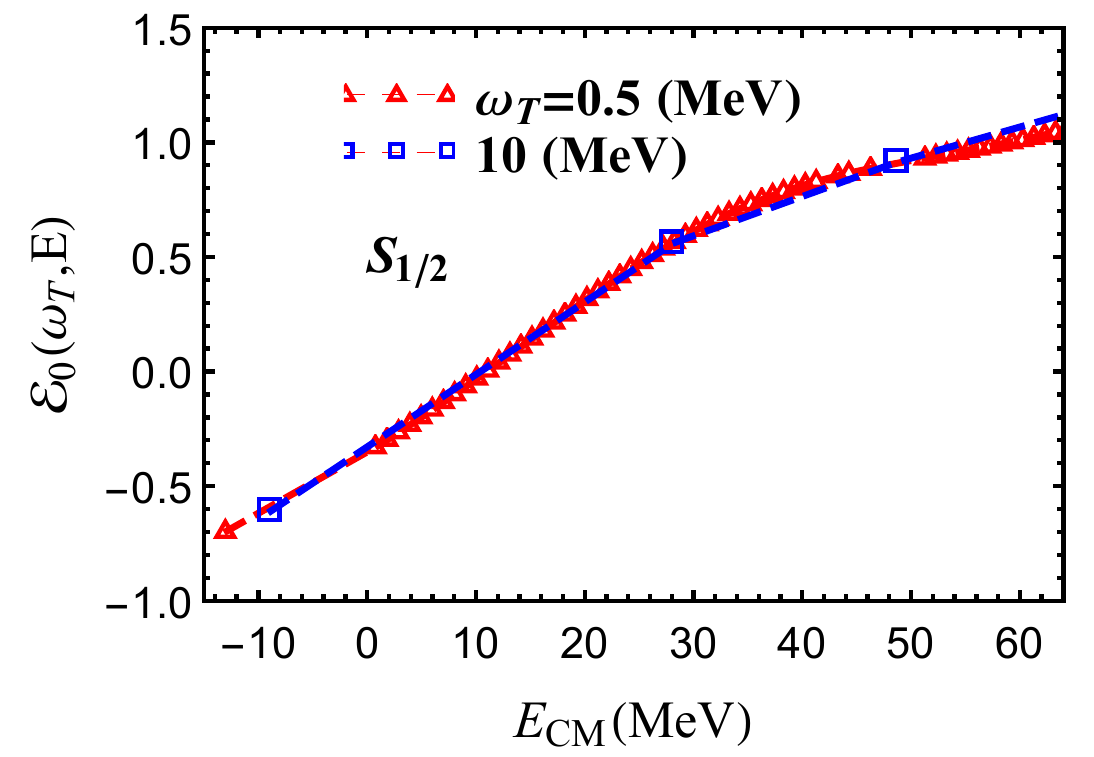}
  \caption{$\gere^\mathrm{exact}_\ell(\omgT, \Eexact)$ values at exact eigenenergies without Hilbert-space truncations for the $\S{1/2}$ channel. } \label{fig:toys1halfgere}
\end{figure}

\begin{figure}[tbh]
  \includegraphics[width=0.45 \textwidth]{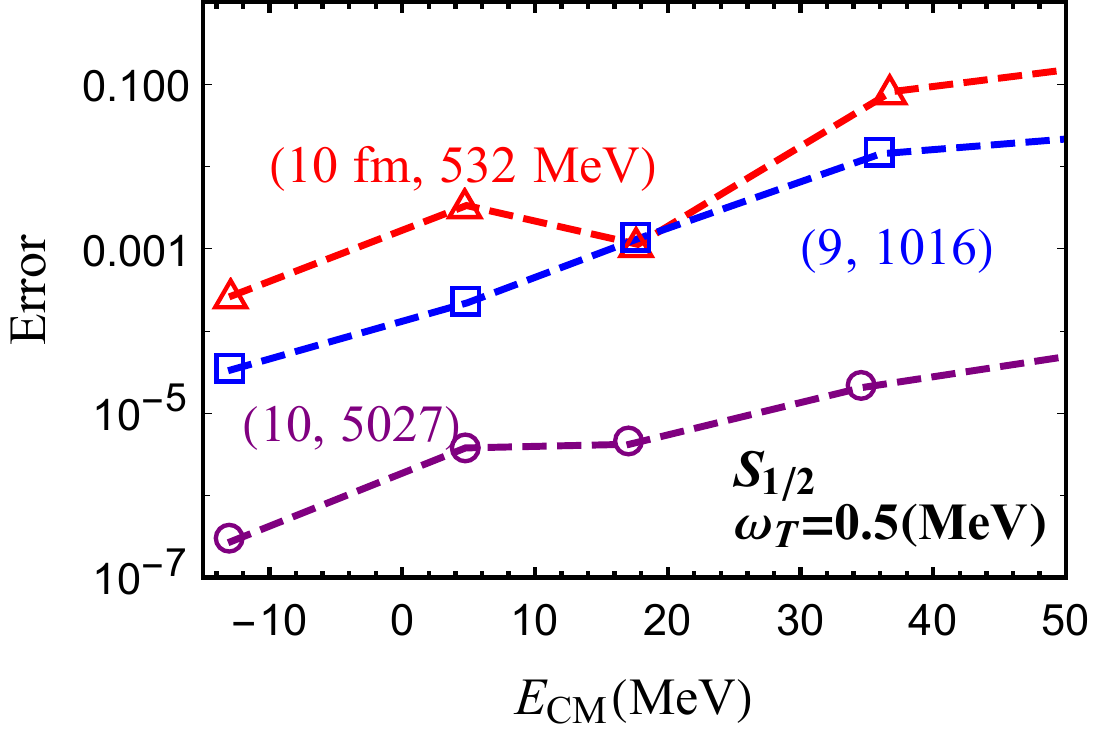}
  \includegraphics[width=0.45 \textwidth]{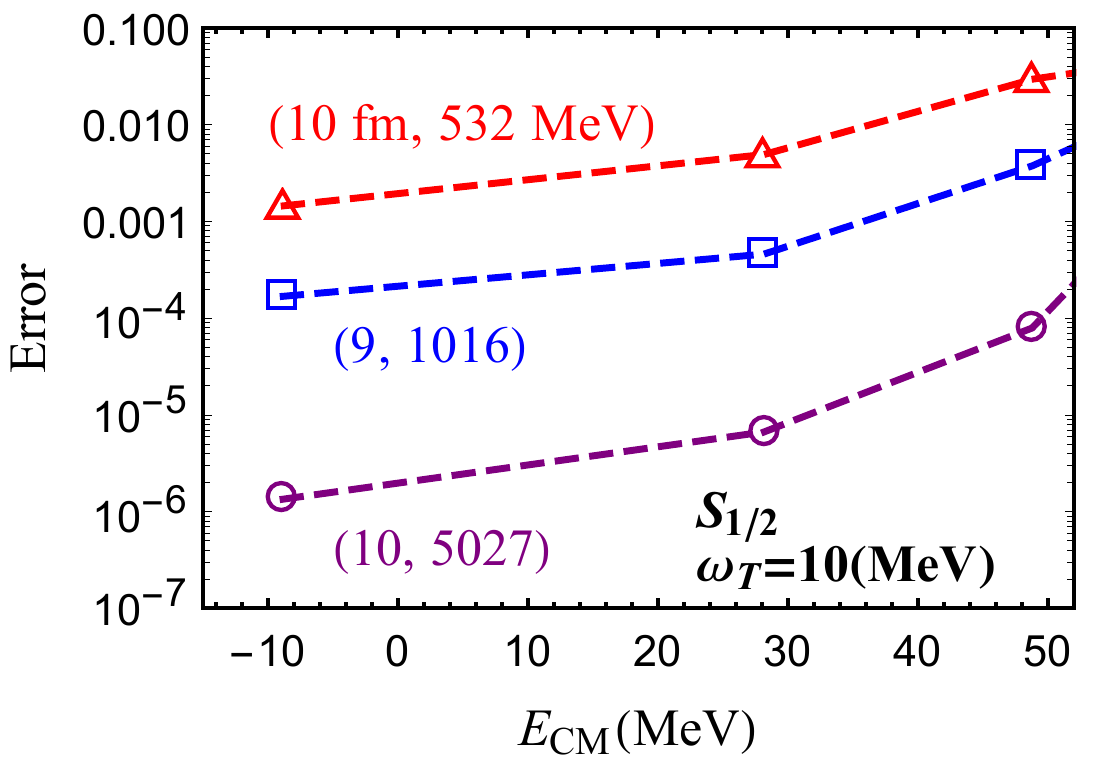}
  \caption{The error of extracted $\gere^\mathrm{regulated}_\ell(\omgT,\Eabi)$ values (i.e., its difference from the exact values $\gere^\mathrm{exact}_\ell(\omgT,\Eabi)$) at eigenenergies of the truncated Hamiltonian for the $\S{1/2}$ channel. The $(\nx ,\omgT)$ values for these calculations are $(10,14)$, $(22,27)$, $(124,132)$ giving increasing $\uvc$.} \label{fig:toys1half}
\end{figure}

As done in Fig.~\ref{fig:toyp3halfgere}, Figs.~\ref{fig:toyp1halfgere} and~\ref{fig:toys1halfgere} plot  $\gere^\mathrm{exact}_\ell(\omgT, \Eexact )$ for the $\P{1/2}$ and $\S{1/2}$ channels and the corresponding $\Eexact$ and $\omgT$ values. 
For the $\P{1/2}$ channel, they show that $\gere^\mathrm{exact}_1(\omgT, E )$ is on the order of $10^{-1}$ when $E$ is below $10$\,MeV, and increases to be on the order of $1$ at higher energies.
For the $\S{1/2}$ channel, $\gere^\mathrm{exact}_0(\omgT, E )$ is in general on the order of $1$.  
This information can be used to infer the magnitude of the relative errors from the absolute errors plotted in Figs.~\ref{fig:toyp1half} and~\ref{fig:toys1half}. 
The latter two plots parallel the top panels in  Figs.~\ref{fig:toyp3half} and~\ref{fig:toyp3half2} but for the $\P{1/2}$ and $\S{1/2}$ channels. The error plots again show systematic improvement of extracted $\gere_\ell$ values with increasing $\uvc$. 

\clearpage 

\section{Data tables and Bayesian-inference-based analysis}

The values of the regulator parameters and trap frequency $\omgT$ used in the data analysis are provided in  Tables~\ref{tab:He-NCSM-p3-2}--\ref{tab:O-IMSRG-d3-2}.
The labeling of the different bins in Figs.~\ref{fig:nHe4} and~\ref{fig:nO24} in the main text are based on the typical $\uvc$ values in those bins (the first column in those tables).  The computed states for the full p-t system and those for the target t are listed in the table captions. 
The detailed data can be found in the  ``Ab\_initio\_energy\_output.zip'' file included in the supplemental material, in which there are four directories: ``Results\_NCSM\_He'',  ``Results\_IMSRG\_He'', ``Results\_IMSRG\_Oxy'', and ``Results\_NCSMC\_He'' (the results from the direct phase-shift calculations). Their names and the names of the files under them  are self-explanatory. Necessary information for understanding the data files are also included therein. It is worth emphasizing that (1) all the energies in the data files have CM energies subtracted and have MeV as units; (2) in our analysis, we have only included the data with $E\leq E_H$ with $E_H \equiv \MH^2/2\mr$ as the high-energy scale in the GERE expansion.

In the following, we elaborate on our data analysis. 	
We start with modeling the errors caused by truncating the many-body Hilbert space.  In order to infer the relative eigenenergy $E$ from the computed eigenenergies $E_{pt}$ and $E_t$, we need to do the following subtraction, as alluded in \eq{eqn:deltatidlenmaxdef},
\begin{eqnarray}
  E(\nx^{pt}, \omg, \omgT) & = &
  E_{pt}(\nx^{pt}, \omg, \omgT)- E_{t}(\nx^{t}, \omg, \omgT) \notag \\ 
  \mathrm{with} \ \nx^{t}  &
  \equiv &  \nx^{pt} + \tilde{\delta}_0  +  \tilde{\Delta}(\nx^{pt},\omg,\omgT)    \;. \label{eqn:deltatidlenmaxdef2}
\end{eqnarray}
The function $ \tilde{\Delta}(\nx^{pt},\omg,\omgT) $ is \textit{a priori}  unknown.  
What is also unknown is the relation between $\nbar$ used in $\ufunc_{\ell}$ and the many-body regulator $\nx^{pt}$ (though $\omg$ and $\omgT$ should be the same). We parameterize this relation as 
\begin{equation}
  2 \nbar + \ell    = 
  \nx^{pt} + \delta_0  +  \Delta(\nx^{pt},\omg,\omgT)   
  \;, \label{eqn:deltanmaxdef}   
\end{equation}
To proceed, we linearize the unknown $\Delta$ function about $\omgT$:  
\begin{equation}
\Delta({\nx^{pt},\omg},\omgT)  = \delta_{\vec{i}}^{(0)} + \delta_{\vec{i}}^{(1)} \times  (\omgT/\omgT^\mathrm{ref}-1)  \ .  \label{eqn:offsets}
\end{equation}
The $\vec{i}$ index denotes $(\nx^{pt},\omg)$ and runs through all the values existing in the p-t NCSM or IMSRG data; for $\omgT^\mathrm{ref}$ we use the mean value of the $\omgT$ values in the output data. 
In the same way, $\tilde{\Delta}$ is linearized about $\omgT$ with $\tilde{\delta}_{\vec{i}}^{(0,1)}$ as unknown parameters. 
Because of the convention of the $\nx$ definition, for the $n$--$\alpha$ p-waves (s-wave), $ \delta_0 =1$ and $\tilde{\delta}_0 = 0$ ($ \delta_0 = 2$ and $\tilde{\delta}_0 = 1$). 
For the IMSRG analysis, the same parameterizations are applied with $\nx \rightarrow \ex$, and different $\delta_{\vec{i}}^{(0,1)}$ and $\tilde{\delta}_{\vec{i}}^{(0,1)}$, but with $\delta_0=\tilde{\delta}_0 = 0 $ for $n$--$\alpha$ and $n$--$\oxy{24}$. 

$C_{i,j,k}$ (labeled $\vec{C}$) and $\qref$, and the other parameters (labeled $\nuipara$) . 

As also mentioned in the main text, the $C_{i,j,k}$ and $\qref$ parameters in \eq{eqn:lambdaUVintp} and the $\nuipara$ parameters (i.e., $\delta_{\vec{i}}^{(0),(1)}$ and $\tilde{\delta}_{\vec{i}}^{(0),(1)}$ with $\vec{i}$ running through the regulator parameters in the given data bin) are inferred by plugging the \abi eigenenergy vector (named as $\Eabi$) into the right side of  \eq{eqn:master2} as ``measured'' $\gere_\ell$ values, and using the GERE expansion of $\gere_\ell (\uvc;\omgT,E)$ as theory. 
In the following discussion, $\gere_\ell$ and $C_{i,j,k}$ are rescaled by $\MH^{2\ell+1}$ and $\MH^{2\ell+1-4i-2j}$, respectively, and become dimensionless. The GERE expansion is truncated (and named as $\gereT_\ell$) by keeping only $C_{i,j,k}$ having $2i+j \leq \geretrun $ (denoted as $\CL$); the contribution of the other terms (labeled as $\CH$) is considered as a series truncation error. Bayesian inference~\cite{Sivia1996, Furnstahl:2014xsa,Furnstahl:2015rha} is used to obtain the joint probability distribution function (PDF) for $\CL$, $\qref$, and $\nuipara$: 
\begin{widetext}
\begin{eqnarray}
 \pr(\CL,\qref,\nuipara\given \Eabi,I) & = & 
   \int  d\Eexact\, d\CH \pr(\vec{C},\qref,\nuipara\given \Eexact,I) \pr(\Eexact\given \Eabi, I)   
  \notag \\ 
  & = &  \int  d\Eexact\, d\CH \pr(\Eexact \given  \vec{C},\qref, \nuipara, I)\pr( \vec{C},\qref, \nuipara  \given  I)   
  \frac{\pr(\Eexact \given \Eabi  , I)}{\pr(\Eexact\given I)}  \ .    \label{eqn:baye1}
\end{eqnarray}
\end{widetext}
Here $\vec{C} = \CL \oplus \CH$, i.e., including all the coefficients in the GREE expansion. 
This joint PDF then  gives $\pr(\CL,\qref\given \Eabi,I)$ after $\nuipara$ is integrated out. 
	
To deal with stochastic numerical errors in the \abi eigenenergies $\Eabi$, \eq{eqn:baye1} is integrated over the exact-energy variable $\Eexact$ with $\pr(\Eexact\given \Eabi, I)$ taken to be an uncorrelated Gaussian distribution (GD) centered at $\Eabi$ with width $0.1$ and $1$ keV for NCSM and IMSRG (the size of their stochastic numerical errors).  
The prior $\pr(\Eexact\given I)$ is a uniform distribution (UD) across a wide energy range (results are not sensitive to the UD's range provided it is on the scale of, or much larger than, $E_H$). 
The prior $\pr( \vec{C}, \qref,  \nuipara  \given  I)$ is separable, with  $\pr(\vec{C}\given I)$ a multivariate GD  centered at $\vec{0}$ and with an identity covariance matrix, $\pr(\qref\given I)$ a UD with $0<\qref<500$\,MeV, and $\pr(\nuipara\given I)$ consists of UDs with $|\delta_{\vec{i}}^{(0)}|$ and $|\tilde{\delta}_{\vec{i}}^{(0)}|$ below $\nx^{pt}/2$ ($\ex^{pt}/2$), and UDs with  $|\delta_{\vec{i}}^{(1)}|$ and $|\tilde{\delta}_{\vec{i}}^{(1)}|$ below 10. Further constraints are $\Delta$ and $\tilde{\Delta}$ being negative and $C_{0,0}(\uvc)$ and $C_{0,1}(\uvc)$ satisfying a causality constraint [see Eq.~(60) and~(61) in~\cite{Hammer:2010fw}]. We set their interaction range parameter $R=5$ and $8$ fm for $n$--$\alpha$ and $n$--$\oxy{24}$, respectively. Note that the causality constraint plays a negligible role in most data bin analyses, except in the resonant channels at lowest $\uvc$ bin. 

In \eq{eqn:baye1}, $\CH$ can be analytically integrated out, resulting in a theory-error covariance matrix for constructing the likelihood function $\pr(\Eexact \given  \CL,\qref, \nuipara, I)$ (c.f.\ Ref.~\cite{Wesolowski:2018lzj}). We then apply the PTEMCEE package~\cite{2013PASP..125..306F,2016MNRAS.455.1919V} (a Markov chain Monte-Carlo sampler implementing parallel tempering) to  sample $\pr(\CL,\qref,\nuipara\given \Eabi,I)$. It is then used to compute error bars for $\gereT_\ell(\uvc;\omgT,E)$. With large $\geretrun$, the series-truncation errors at the data points become much less than the $\nuipara$-induced errors, and $\CH$ would not be constrained by data but only by its prior~\cite{Wesolowski:2015fqa}. The series truncation error for $\gere_\ell$ at given $\uvc$, $\omgT$, and $E$, is then an infinite sum of GDs (with $\qref$ at its mean value), i.e., also a GD with zero mean and a simple variance ($\sigma^2_{\gere, \mathrm{th}}$); its correlation with $\gereT_\ell$ becomes
negligible. (Also note that the series truncation error would grow out of control when $E$ increases beyond $E_H$. Therefore we only include the data with $E\leq E_H$ in our analysis.)
As the result, to compute error bars for $\gere_\ell$, the error bars for the truncated $\gereT_\ell$, can be  added in quadrature with the series truncation error $\sigma_{\gere, \mathrm{th}}$. 
In this work, we compromise between reducing that correlation and the numerical effort, so we use $\geretrun=12$ for analyzing NCSM $n$--$\alpha$ and IMSRG $n$--$\oxy{24}$, but $\geretrun=6$ for IMSRG $n$--$\alpha$.
Our Bayesian inference formalism is discussed in more detail in Ref.~\cite{JordanMelendezPhD}.

\begin{table}
\caption{Regulator parameter values and $\omgT$ values for different data bins for the NCSM $n$--$\alpha$ $\P{3/2}$ channel. The involved states are  $\alpha$'s ground state ($0^+$) and $\he{5}$'s $3/2^-$ computed with the listed regulator and $\omgT$.  } \label{tab:He-NCSM-p3-2}
\begin{ruledtabular} 
   \begin{tabular}{cccc}
 $\uvc$ (MeV) & $\nx$ & $\omg$ (MeV) & $\omgT$ (MeV)        \\ \hline  
$\sim 900$ &  12,14,16 & 28 &  2,4,6,8,10,12,14,16 \\ \hline 
$800$ &  8,10 &  28 &  2,4,6,8,10,12,14,16 \\
      &  14,16 &  20 &  4,6,8,10  \\ \hline
$700$ &  10,12 & 20 &  4,6,8,10 \\ 
      &  14,16 & 15 &  4,6,8,10  \\       
\end{tabular} 
	\end{ruledtabular}
	
\caption{Regulator parameter values and $\omgT$ values for different data bins for the NCSM $n$--$\alpha$ $\P{1/2}$ channel. The involved states are  $\alpha$'s ground state ($0^+$) and $\he{5}$'s $1/2^-$ computed with the listed regulator and $\omgT$.  } \label{tab:He-NCSM-p1-2}
\begin{ruledtabular} 
   \begin{tabular}{cccc}
 $\uvc$ (MeV) & $\nx$ & $\omg$ (MeV) & $\omgT$ (MeV)        \\ \hline  
$\sim 900$ &  12,14,16 & 28 &  2,4,6,8,10,12 \\ \hline 
$800$ &  8 &  28 &  2,4,6,8,10 \\
      &  10 &  28 &  2,4,6,8,10,12 \\
      &  14,16 &  20 &  4,6,8,10  \\ \hline
$700$ &  10,12 & 20 &  4,6,8,10 \\ 
      &  14,16 & 15 &  4,6,8,10  \\       
\end{tabular} 
	\end{ruledtabular}
	
\caption{Regulator parameter values and $\omgT$ values for different data bins for the NCSM $n$--$\alpha$ $\S{1/2}$ channel. The involved states are  $\alpha$'s ground state ($0^+$) and $\he{5}$'s $1/2^+$ computed with the listed regulator and $\omgT$. } \label{tab:He-NCSM-s1-2}
\begin{ruledtabular} 
   \begin{tabular}{cccc}
 $\uvc$ (MeV) & $\nx$ & $\omg$ (MeV) & $\omgT$ (MeV)        \\ \hline  
$\sim 900$ &  11, 13, 15 & 28  &  2,4,6,8 \\ \hline 
$750$ &  9 & 28  &    2,4,6 \\ 
      &  13,15 & 20  & 4,6,8  \\   \hline 
$650$ &  9, 11 & 20  &  4,6 \\ 
      &  13, 15 & 15  & 4,6,8 \\
\end{tabular} 
\end{ruledtabular}
	\end{table}	
 \begin{table}
\caption{Regulator parameter values for different data bins for the IMSRG $n$--$\alpha$ $\P{3/2}$ channel. Note $\omgT =  2-10,12,14,16$\,MeV for all the data bins. The involved states are  $\alpha$'s ground state ($0^+$) and $\he{5}$'s $3/2^-$ computed with the listed regulator and $\omgT$.  } \label{tab:He-IMSRG-p3-2}
\begin{ruledtabular} 
   \begin{tabular}{ccc}
 $\uvc$ (MeV) & $\ex$ & $\omg$ (MeV) \\ \hline  
$\sim 900$ &  12,14 & 28   \\
      &     14 & 24  \\ \hline 
$800$ & 10 & 28     \\
      & 12 & 24     \\ 
      & 14 & 20   \\  \hline 
$750$ & 10 & 24     \\
      & 12 & 20     \\  
      & 14 & 16      \\ \hline     
$650$ & 10 & 20    \\
      & 10,12& 16  \\ \hline 
$550$ & 10,12,14 & 12   \\
\end{tabular} 	\end{ruledtabular}

\caption{Regulator parameter values for different data bins for the IMSRG $n$--$\alpha$ $\P{1/2}$ channel. Note $\omgT =  2$--$10$\,MeV for all the data bins. The involved states are  $\alpha$'s ground state ($0^+$) and $\he{5}$'s $1/2^-$ computed with the listed regulator and $\omgT$. } \label{tab:He-IMSRG-p1-2}
\begin{ruledtabular} 
   \begin{tabular}{ccc}
 $\uvc$ (MeV) & $\ex$ & $\omg$ (MeV)         \\ \hline  
$\sim 900$ &  12,14 & 28   \\
      &     14 & 24  \\ \hline 
$750$ & 10 & 28      \\
      & 10,12 & 24   \\ 
      & 14 & 20     \\  \hline 
$700$  & 10,12 & 20      \\  
      & 12, 14 & 16       \\ \hline     
$550$ & 10 & 16    \\
      & 10,12,14 & 12   \\     
\end{tabular} 	\end{ruledtabular}

\caption{Regulator parameter values for different data bins for IMSRG $n$--$\oxy{24}$ $\D{3/2}$ channel. $\omgT = 1,1.5,2,2.5,3,3.5,4$ for all the data bins. The involved states are  $\oxy{24}$'s ground state ($0^+$) and $\oxy{25}$'s $3/2^+$ computed with the listed regulator and $\omgT$. } \label{tab:O-IMSRG-d3-2}
\begin{ruledtabular} 
   \begin{tabular}{ccc}
 $\uvc$ (MeV) & $\ex$ & $\omg$ (MeV)       \\ \hline  
$\sim 900$ &  14 & 28    \\
      &     14 & 24    \\ \hline 
$700$ & 14 & 20      \\
      & 14 & 16     \\  \hline 
$800$  & 14 & 24    \\  
      & 14 &  20   \\ 
\end{tabular} 	\end{ruledtabular}
	\end{table}

\end{document}